\documentclass[12pt]{article}
\usepackage[english]{babel}
\usepackage[letterpaper,top=2cm,bottom=2cm,left=3cm,right=3cm,marginparwidth=1.75cm]{geometry}
\usepackage{booktabs}
\usepackage{amsmath}
\usepackage[colorlinks=true, allcolors=blue]{hyperref}
\usepackage{lmodern}
\usepackage[T1]{fontenc}
\usepackage{graphicx}
\usepackage{verbatim}
\usepackage{afterpage}
\usepackage{amsfonts}
\usepackage{amsthm}
\usepackage{amssymb}
\usepackage{setspace}
\usepackage[round]{natbib}
\usepackage{pdfpages}
\usepackage{subcaption}
\usepackage{authblk}
\usepackage{array}
\usepackage{dcolumn}
\usepackage{longtable,lscape}
\usepackage{rotating}
\usepackage{pdflscape}
\usepackage{todonotes}
\usepackage[para]{threeparttable}
\usepackage{tabularx}
\usepackage{marginnote}
\usepackage{xfrac}
\usepackage[normalem]{ulem}
\usepackage{rotfloat}
\usepackage{float} 

\usepackage{footmisc}
\usepackage{adjustbox}

\usepackage{url}

\def\sym#1{\ifmmode^{#1}\else\(^{#1}\)\fi}

\makeatletter
\def\thanks#1{\protected@xdef\@thanks{\@thanks
        \protect\footnotetext{#1}}}
\makeatother

\begin{document}

\title{
\vspace{-20pt}

%Hobo Economicus Rides Again
%Poor but Rational
%Down and Out among the Econs
%No Income Targets in the Sale of Street Papers
%Hobo Economicus Sells Street Papers
%Labor Supply under Temporary Wage Increases: Evidence from a Randomized Field Experiment
Labor Supply among Homeless Street-Paper Sellers: Evidence from a Randomized Wage Increase

\thanks{Ekman: Karlstad University, Karlstad Business School, 651 88 Karlstad, Sweden (e-mail: mats.ekman@kau.se). 
Jakobsson: Karlstad University, Karlstad Business School, 651 88 Karlstad, Sweden, and FBK-IRVAPP, Trento, Italy (e-mail: niklas.jakobsson@kau.se). 
Kotsadam: Ragnar Frisch Centre for Economic Research, Gaustadalleen 21, 0349 Oslo, Norway (e-mail: andreas.kotsadam@frisch.uio.no). We are grateful to the staff at \textit{Situation Sthlm} for their smooth and efficient collaboration, and gratefully acknowledge the financial assistance of the Jan Wallander and Tom Hedelius Foundation, and Tore Browaldh’s Foundation (Grant no. P23-0086) and the Norwegian Research Council (project number 341250). We are grateful for comments and suggestions from Oddbjørn Raaum, Siri Jakobsson Store, Carl Bonander, and several seminar and conference participants. Refine.ink was used to check the paper for consistency and clarity. An analysis plan \citep{preanalysisplan} is pre-registered at The American Economic Association’s registry for randomized controlled trials (No. AEARCTR-0014308), and all deviations from the plan are noted in the text. The plan is also found in the Appendix. This study was reviewed by the Karlstad University Ethical Review Board and the Swedish Ethical Review Authority and was confirmed exempt from the Swedish Ethical Review Act (Dnr 2023-05373-01). Before randomization, all participating sellers provided informed consent. Participation was voluntary, and sellers could withdraw at any time without consequences. The study was designed and conducted in close collaboration with the editorial management of Situation Stockholm, with careful attention to the welfare of participating sellers.}
}

\author{Mats Ekman, Niklas Jakobsson, and Andreas Kotsadam}

\maketitle
\vspace{-2em}  % adjust -2em as needed (try -1em, -1.5em, etc.)
\thispagestyle{empty}

\begin{abstract}
\noindent 
We conduct a pre-registered randomized controlled trial to test for income targeting in labor-supply decisions among sellers of a Swedish street paper. Unlike most workers, these sellers choose their own hours and often face severe liquidity constraints and volatile incomes. Treated individuals received a 25 percent bonus per copy sold via Swish, the mobile payment system that accounts for roughly 80 percent of sales, for the duration of an issue, simulating an increase in earnings potential. Consistent with standard labor-supply theory, they sold more papers and, by our measures, worked longer hours and took fewer days off. These findings contrast with studies on intertemporal labor supply that find small substitution effects.
\small

\vspace{2ex}

\newline

\newline

\end{abstract}

\clearpage
\doublespacing

\section{Introduction}
Standard labor supply theory accommodates less work at higher pay by reference to
the possibility of a backward-bending supply curve, but when the increase is temporary, standard theory predicts more work to enable both more money
and leisure to be had over time \citep[e.g.,][]{friedman1957theory, modigliani1954utility}. The compensated response is unambiguously positive, and the wealth effect that nets against it is second-order for a transitory shock. Models of reference-dependent preferences \citep{Kahneman1979prospect} instead posit high marginal costs of effort once an income target is reached, so that workers choose more leisure during positive transitory income shocks (modulated in various incarnations; e.g., \citet{koszegi2006model} model targets as expectations). Causal evidence from experimental variation on this question remains scarce, particularly for workers in precarious economic and social circumstances. Low and volatile earnings, together with liquidity constraints, make this a consequential setting for studying responses to temporary changes in earnings potential. It is also worth looking for reference dependence among such workers if one believes that a failure to maximize contributes to their stark lack of worldly success. At the same time, standard theory predicts large substitution effects when baseline earnings are low. We conduct a randomized controlled trial to test for income targeting in labor-supply decisions among mostly homeless sellers of a Swedish street paper. To the best of our knowledge, this paper is the first to test for income targeting among this kind of worker.

There has been growing interest in studying workers with uncertain incomes who set their own hours. Pioneering work by \citet{camerer1997labor} finds strong evidence of reference-dependent behavior among New York City cab drivers, using observational data. Since then, the taxi industry has been studied further by \citet{farber2005tomorrow, farber2008reference, farber2015taxi}, \citet{crawford2011newyork}, \citet{agarwal2015singaporean}, \citet{martin2017when}, \citet{thakral2021daily}, and \citet{duong2023taxi}, among others. Some of these findings challenge the original conclusions, while others clarify the conditions under which income targeting is most likely to occur, such as specific income thresholds or planning horizons. In light of these developments, it is particularly interesting to study workers whose personal characteristics may make maximization less expected. The fact that our subjects do much less well than do people around them suggests that reference dependence should be particularly easy to find among them, whereas the classical framework predicts a strongly dominant substitution effect from greater earnings potential, given that these workers earn so little and plausibly have few alternative sources of income.

A small number of randomized experiments address labor supply responses to transitory changes in earnings \citep{fehrgoette,andersen2025toward,dupas2020daily,angrist2021uber}, generally finding positive elasticities, although interpretations differ. \citet{fehrgoette} increased the earnings potential of Swiss bike messengers and found a positive labor-supply response but a negative elasticity of effort per hour. Offering their subjects the chance to participate in lotteries and observing their decisions, they conclude that reference dependence best fits the data. In a two-pronged setup, \citet{andersen2025toward} paid vendors in an Indian open-air market a supplemental hourly wage, and gave vendors an unexpected windfall through an overpayment by a confederate. In the first experiment, treated vendors worked longer hours, although their response came with a one-day lag; in the second, hours were unaffected. \citet{dupas2020daily} find that Kenyan bicycle-taxi drivers' labor supply is driven by daily cash needs. In an experimental study of Uber drivers, \citet{angrist2021uber} find positive labor-supply elasticities in response to randomized variation in commission rates, which they read as evidence against income targeting; the reluctance of their drivers to take up a more profitable fixed-lease contract they attribute to loss aversion over the fixed cost rather than to a target.

Our subjects are sellers of a major Swedish street paper, \textit{Situation Sthlm} (short for ‘Stockholm’). Street papers are sold by homeless or precariously housed individuals, typically as a way for them to come into or remain in gainful employment.\footnote{In 1915, \textit{Hobo News} was the first such paper, published by the Midwest-based International Brotherhood Welfare Association, and current examples include \textit{The Big Issue} (the UK), \textit{StreetWise} (Chicago), and \textit{StreetSheet} (San Francisco).} The sellers are paid solely on commission, and although we cannot know for sure, the street paper is likely a large share of total earned income for at least the more eager among the sellers in our sample, for whom sales bring in roughly 3,400 to 4,500 SEK a month without the bonus (see estimates at 40 SEK per paper in Tables \ref{tab:main}, \ref{tab:summary}, and \ref{tab:waitlist} below). Since these individuals are homeless or precariously housed and eligible for social assistance, alternative income sources are unlikely to account for much of their effort.

To mimic favorable business conditions, we give treated sellers a bonus of 10 SEK (about 1 US dollar) per copy sold through the common Swedish electronic payment system Swish. Sellers otherwise earn 40 SEK per copy, so our intervention is a 25 percent bonus. We pay the bonus on sales of the October issue, from the day of its release until the release of the November issue. 

Treated sellers sold more papers. Measures of working time and days worked, constructed from Swish-transaction timestamps, move in the same direction but are mechanically sensitive to the bonus's payment mode, and we treat them as suggestive. Because the intervention bundles a piece-rate change with payment-mode, liquidity, and salience effects, we read the experimental result primarily as evidence against income targeting in this population rather than as a clean estimate of a wage elasticity.  

Applying an observational strategy similar to the taxi-driver literature to pre-treat\-ment sales data yields a large and negative hours-wage elasticity that, taken at face value, would suggest income targeting. This estimate is substantially inflated by division bias: weather-instrumented estimates attenuate it toward zero, and a complementary test asking whether high daily earnings reduce next-day labor supply finds the opposite of income targeting (good days predict \emph{more} work the next day, not less). We read these analyses as internal consistency checks rather than as independent identification of the wage elasticity, given weak first stages for the weather instruments and alternative explanations for the next-day pattern.

Our approach differs from most previous analyses in at least three key respects. First, the population is different. Where the literature has studied cabdrivers, bicycle messengers, and platform drivers, labor supply among our subjects carries relatively large welfare implications: additional income affects how they live more than the same additional income does for others.

Second and relatedly, following \citet{leeson2022hobo}, it is particularly useful to test rational-choice predictions in a setting where such predictions are often viewed as least likely to hold. \citet{leeson2022hobo} study where panhandlers decide to locate along the Washington DC Metro System, finding that their time-adjusted earnings tend toward equality (partly, but not fully, reproduced by \citet{demeloetalreplication}). Their approach does not accommodate tests for income targets, but similar priors about the homeless motivate our study. Finding positive labor supply elasticities to transitory income shocks among sellers of street papers is informative about the reach of standard labor-supply models.

Third, our measurement also adds perspective to prior literature on the elasticity of labor supply. Related to the issue of division bias \citep{borjas1980wages}, a risk in some studies focusing on taxi drivers is that the tendency to use trip-level data for individual drivers may conflate leisure and work. Once a driver is on the road, the taxi lease is a sunk cost, and so what may look like work hours may actually be drivers running personal errands, making work hours on slow days appear longer than they actually are. Some studies, such as \citet{duong2023taxi}, use detailed administrative trip data to address the question of when drivers actually work. Such data records when drivers are on duty or carrying passengers, as opposed to when they are unavailable, but drivers may nevertheless be on duty while running personal errands and so not actively engaged in working. Because our outcome is sales during the period when an issue is current, we avoid this specific form of measurement error by relying on an administratively recorded output measure rather than constructing hours from noisy or incidental activity data. However, sales as an output measure inherently reflect both time worked and effort per unit of time, so they do not isolate the hours margin. Under the monotonicity assumption that total sales increase with work time, sales serve as our main proxy for time spent working for both the control and treated groups; treated sellers, in addition, receive higher pay per copy sold during the bonus period. We also use more detailed measures of input (Hours and Days based on electronic timestamps) to address the separate margins of time and effort.

The remainder of this paper proceeds as follows. In Section 2, we present the experimental design and the data set. In Section 3, we present the pre-specified empirical strategy, our findings, heterogeneity analyses, a series of robustness checks, some survey evidence, and observational evidence on reference dependence. Section 4 concludes.

\section{Experimental Design and Data}
\subsection{Experimental Design}
\textit{Situation Sthlm} was founded in 1995 and has an annual circulation of approximately 200,000 copies, sold by about 250 sellers. The sellers are a heterogeneous group, but tend to lack permanent housing and face mental-health and substance-use challenges at higher rates than the general population. Apart from selling the street paper, their sources of income include social assistance, panhandling, and deposits from empty bottles and cans they collect. Some sellers also occasionally write articles or do publicity work for the street paper. As we hypothesize in the Introduction, given their income from sales, it is likely, though not certain, that these alternatives, unobserved by us, do not consume a large share of the more active sellers' total working time. When they sell the paper, they must first purchase it from the office of \textit{Situation Sthlm}, located in Central Stockholm. At the time of the intervention, the price of a copy to sellers was 40 SEK, and the resale price was 80 SEK. The sellers have designated spots across Stockholm where they are encouraged to sell, but they are not restricted to these locations. Unsold copies must be returned by the release of a new issue, and \textit{Situation Sthlm} refunds the entire 40 SEK per copy in such cases. A new issue is released on the last Wednesday of every month except for July (the June issue is a double issue). The sellers may come and go to the office as they please during the office’s hours of operation.

When an electronic sale is made, the payment system allows the seller to obtain the funds from the office of \textit{Situation Sthlm} after about 45 minutes. For severely budget-constrained sellers, this fact implies a cap on daily sales, which is not necessarily raised by our bonus. For example, a seller who has only a hundred SEK could buy two copies of the paper (80 SEK) during the intervention, and come back to the office to obtain 180 SEK with our bonus (enough for five new copies if the seller retained the 20 SEK left over from the first transaction) when these sales have been made and the 45-minute processing of the payment has been concluded. If the seller can sell all five papers but the office closes before they can restock, the bonus will increase this individual's sales by just one copy on that day (five rather than four). If demand falls short of five, the bonus has no effect on sales for that day. During the intervention, the office was open from 8:00 a.m. to 3:15 p.m., Monday to Friday, and from 10:00 a.m. to 3:00 p.m. on Saturdays.

On September 23, 2024, an announcement was posted on a wall in the office lobby of \textit{Situation Sthlm}, informing readers that the October issue of \textit{Situation Sthlm}, to be released on September 25, would include a ``Swish campaign''. Swish is the largest mobile payment system in Sweden, and approximately 80 percent of all copies of \textit{Situation Sthlm} are sold using it. The announcement also said that sellers who wish to participate in the campaign will watch a coin flip, and if the coin comes up heads, they receive an extra 10 SEK per copy of the October issue sold via Swish, released on September 25. If the coin comes up tails, they receive a bonus for the next issue, released on October 30 (the November issue). Hence, the assignment follows a randomized waitlist-control design, with delayed treatment. The coin flips were administered as sellers came into the office to purchase copies of the new issue, starting on its release date. Sellers were thus assigned to a group upon entrance and agreement to participate. A total of 56 sellers were assigned to the October treatment, and 53 sellers to the November treatment. Five sellers chose not to participate. Because control sellers knew their bonus would arrive during the November issue, the counterfactual is not a neutral baseline but rather an announced delayed treatment. We flag two potential concerns upfront. First, this design does not cleanly identify a pure intertemporal substitution response: a rational late-treated seller anticipating a November bonus may shift effort from October to November, which would bias the estimated October treatment effect upward (since controls work less than they otherwise would in October). Second, when we later use the originally-treated sellers as controls in the waitlist analysis (Table~\ref{tab:waitlist}), those ``controls'' are contaminated by potential habit formation and learning from their own October treatment. Additionally, on November 18, the office announced a permanent increase in the cover price to 100 SEK, effective with the December issue released on November 27, with sellers retaining 50 SEK rather than 40 SEK per copy sold. The change applied to all payment modes and raised the sellers' acquisition cost along with the price, so it did not make the bonus itself permanent; what it did was place a second, non-transitory change in earnings potential in front of the late-treated cohort in the middle of their bonus period\footnote{Such an increase had been discussed already the year before, and the editorial office's preferred time to implement it is the December issue to take advantage of higher demand during Christmas. Thus, it is possible that some fraction of the sellers anticipated a price increase before it came into effect.}. We therefore do not interpret the November estimate as a clean short-term causal effect of the bonus, and we treat the waitlist analysis as a robustness check whose interpretation is conditional on assumptions about contamination.

Tying the bonus only to Swish sales was done to prevent sellers from purchasing papers from the office for immediate resale to colleagues randomized not to get the bonus.\footnote{This is also the reason for the departure from letting treated sellers purchase copies at a discount, mentioned in our Pre-analysis plan (see Appendix~\ref{sec:pap}). The bonus comes to the same thing, but avoids the risk of resale.} The paper has a pre-existing system to prevent fraud, in which sellers must provide written consent not to use the payment system for private purposes. Essentially, the office has access to the names and phone numbers of all customers who pay using Swish, as well as the amounts received, so whenever one customer buys several copies, or the amount paid differs from 80 SEK, the seller knows that he or she will have to explain this to the office. \textit{Situation Sthlm} incurs a cost to enable customers to pay using Swish, so, naturally, they should not wish to pay for non-sale transactions.

\subsection{Dataset}
\textit{Data sources:} We use data from the editorial office of \textit{Situation Sthlm} on all sales by individual sellers. Sales are the number of copies sold. For electronic sales, we have access to real-time sales data; for cash payments, we know the issue was sold between when the seller picked it up and when they returned any unsold magazines. In addition to the sales data, we have access to the sellers' ages and sexes. All \textit{Situation Sthlm} sellers in Stockholm who entered the \textit{Situation Sthlm} office during the intervention period were invited to participate in the experiment. A total of 109 sellers agreed to participate, and five declined. A member of staff flipped a coin in front of the participant, thereby assigning 56 sellers to the treatment group and 53 to the control group. Nine sellers in the register were based outside Stockholm and therefore not in direct contact with the \textit{Situation Sthlm} office; they were not asked to participate.

\textit{Variable definitions:} Our main outcome variable is sales of the October issue (Sales, the number of copies sold), which includes all sales made of the issue. Our pre-analysis plan also includes the natural logarithm (+1) of the sales variable (ln(Sales)) and sales via the electronic system only (Electronic), which excludes cash sales, as exploratory outcome variables. Such outcomes also include cash sales only (Cash), which are not directly observed cash transactions but rather a residual constructed by subtracting Electronic sales from Sales, where Sales is inferred from inventory movements and Electronic is measured from Swish transactions. Cash can therefore occasionally take negative values. In our data, Cash has eight negative observations, mainly because some sellers may accept Swish payments without handing over a magazine, or because some buyers pay for several issues while taking only one physical copy, resulting in Swish payments exceeding the number of copies leaving the inventory. In addition, we know the number of issues purchased by the sellers from the editorial office, which we label intended sales. We also examine time spent working as an exploratory outcome variable. Hours worked during the month are approximated by the span between the first and last electronic sale each day (Hours). The number of working days is also approximated by the total number of days in which a seller sells an issue via the electronic system (Days). In common with some taxi studies on supply elasticity \citep[e.g.,][]{camerer1997labor,farber2005tomorrow,crawford2011newyork}, we cannot tell whether a day without sales is a working day (an unsuccessful one). We therefore use a daily binary indicator for observed active sales days, which is summed within issue cycles to construct the monthly Days count. Because Hours and Days are both constructed from electronic transactions, they may be mechanically affected by the bonus, which specifically incentivizes Swish use: a treated seller might ring up a Swish sale earlier or later in the day than a similarly busy untreated seller, widening the measured span without actually working longer. We address this concern in two ways below: by reporting effects within the subsample of sellers who already sell mostly via Swish (Appendix Table~\ref{tab:swishshare}), and by reporting alternative output-based effort measures (sales per active day and sales per hour), which ask whether output rose by more than the measured work window (Appendix Table~\ref{tab:alt_hours}); these ratios are not free of transaction timing either, since Hours and Days enter their denominators. Lastly, the control variables for Age (Age) and sex (Woman) are reported to us by \textit{Situation Sthlm}, and are based on the personal identity numbers of the sellers (these are the Swedish national identification numbers, which are structured to reveal their holders’ sex and year of birth).

\textit{Summary statistics:} Appendix Table \ref{tab:summary} presents summary statistics for the key variables in the study. The average age of the participants is 55.8 years (SD = 10.87), with a range of 26 to 82 years. About 34 percent of the sample are women. The mean number of magazines sold during the October issue is 112.85 (SD = 191.07), with considerable variation, ranging from 0 to 1,555. The natural logarithm of sales (ln(Sales)) is also reported, with an average of 3.61 (SD = 1.79). Sales are further disaggregated into electronic and cash sales. Electronic sales constitute the majority, with a mean of 96.36 (SD = 177.08) and a maximum of 1,467. Cash sales are lower on average, at 16.50 (SD = 25.38), with values ranging from -10 to 160. As indicated above, Cash sales include a few negative observations, mainly because some sellers may sell a magazine to a customer without handing it over, and some buyers may also buy several magazines while taking only one. The participants worked an average of 30 hours per month (SD = 42.68), with some individuals not working at all (e.g., only making cash sales) and others working up to 286.7 hours. The number of days worked during the October issue averages 11.17 (SD = 9.55), with a range from 0 to 33 days.

\textit{Randomization tests:} To assess the validity of the randomization, we test for balance in the control variables between treated and controls. Appendix Table \ref{tab:balance} presents baseline means for treated (n = 56) and controls (n = 53), their differences, and p-values after t-tests. Few differences are statistically significant, but sales in the control group were consistently higher the year before treatment. One further imbalance concerns activity in the year before the experiment: 30.4 percent of treated sellers but only 13.2 percent of controls had no sales at all in the two issues released one year before the intervention ($p = 0.03$). Baseline activity, not only its level, therefore differs across arms, which is a further reason to control for pre-treatment sales throughout. As a joint test of balance, we regress the treatment dummy on the control variables, indicating imbalance at baseline (F = 7.16, p < 0.01), implying that the baseline variables jointly predict treatment status. We therefore control for pre-treatment sales behavior in all analyses to address this imbalance. Including lagged outcomes also improves precision, as prior values are strongly correlated with the outcome (see Appendix Table \ref{tab:correlations}).

\section{Results}
\subsection{Pre-specified regression model}
We estimate several versions of equation~\eqref{eq:regression}:

\begin{equation}
    Y_i = \alpha + \beta_1 T_i + X_i'\beta_2 + \varepsilon_i,
    \label{eq:regression}
\end{equation}

where $Y_i$ is an outcome for seller $i$, $T_i$ is an indicator of treatment, and $X_i$ is a set of baseline controls, including age, sex, and 12 lags of the outcome variable. The regressions are estimated with heteroscedasticity-robust standard errors.

We use the post-double-LASSO selection approach of \citet{belloni2014inference}. The LASSO selection approach selects variables correlated with treatment and outcomes, improving the precision of the estimates and helping correct imbalances across groups. This is especially important since sales in the treated group were consistently lower in the year leading up to treatment.

\subsection{Main findings}
In Table \ref{tab:main}, we present results for the overall population of sellers using equation~\eqref{eq:regression} and the post-double LASSO selection approach. The coefficient of our main outcome variable, Sales, is 23.25, and the mean sales in the control group is 110.68, implying a 21 percent increase in copies sold following treatment. For ln(Sales), the effect is positive but less precisely estimated. Based on the coefficients for Electronic sales and Cash sales, the increase is driven by Electronic sales, as expected. The coefficient on Hours is 10.99 (control group mean 27.11), and on Days, 2.56 (control group mean 10.66), implying large effects on working time. As a simple income-targeting benchmark, if all sales were bonus-eligible and sellers held their monthly income fixed, a 25 percent increase in commission would imply a 20 percent decrease in the number of copies sold. Instead, we find evidence of sales and sales efforts increasing.

Our implied labor-supply elasticities are large, though best interpreted as approximate incentive elasticities rather than structural wage elasticities, given that the bonus applies only to Swish sales and our hours measure is noisy. If the relevant quantity is sales, the increase from 110.68 to 133.93 copies represents a 21.0 percent increase, corresponding to a 25 percent increase in the per-copy return on Swish sales, or an elasticity of approximately 0.8. Our measure of hours indicates a 40.5 percent increase; using the piece-rate shock as the relevant wage change yields an implied hours elasticity of roughly 1.6. Days worked increase by 24.0 percent, implying an elasticity close to 1 if benchmarked against the 25 percent bonus. We do not put weight on elasticities constructed from realized hourly earnings, since realized hourly pay combines the experimental incentive with endogenous changes in sales per hour and the mix of Swish and cash sales. Our measure of hours is more likely a lower bound for the control group and an upper bound for the treated group, given the possibility of extended breaks. The number of hours among our subjects is also low compared to most intensive-margin estimates of supply elasticities for the general public, who tend to work full-time, which are around 0.2 at most \citep{Elminejad2023intertempsub, Eversetal2008WageElast, McClellandMokRev2012}. The elasticity of taxable income with respect to the net-of-tax rate, a related public-finance benchmark that captures both income and substitution channels, similarly tends to lie below 0.4 in the literature. The much larger responses we estimate here are consistent with a substitution effect dominating a small income effect in a population with limited alternative income channels, but they are also large relative to standard estimates. We discuss alternative interpretations (liquidity relief, salience, perceived monitoring) in the conclusion.

\begin{table}[H]
\centering
\begin{threeparttable}
\caption{\label{tab:main} Main results}
\small\setstretch{1}
\begin{tabular}{l*{6}{l}}
\toprule
            &\multicolumn{1}{c}{(1)}&\multicolumn{1}{c}{(2)}&\multicolumn{1}{c}{(3)}&\multicolumn{1}{c}{(4)}&\multicolumn{1}{c}{(5)}&\multicolumn{1}{c}{(6)}\\
            &\multicolumn{1}{c}{Sales}&\multicolumn{1}{c}{ln(Sales)}&\multicolumn{1}{c}{Electronic}&\multicolumn{1}{c}{Cash}&\multicolumn{1}{c}{Hours}&\multicolumn{1}{c}{Days}\\
\midrule
Treated     &       23.25** &        0.30   &       20.14** &        0.95   &       10.99***&        2.56** \\
            &      (9.33)   &      (0.21)   &      (8.93)   &      (3.40)   &      (3.45)   &      (1.06)   \\
\midrule
Mean for controls&      110.68   &        3.54   &       93.57   &       17.11   &       27.11   &       10.66   \\
N           &         109   &         109   &         109   &         109   &         109   &         109   \\
\bottomrule
\end{tabular}

\begin{tablenotes}[flushleft]
\small
\item \textit{Notes:} Treatment effects estimated using the post-double LASSO selection approach. Each column corresponds to a different outcome variable. N = 109. Robust standard errors in parentheses. \sym{*} \(p<0.10\), \sym{**} \(p<0.05\), \sym{***} \(p<0.01\).
\end{tablenotes}
\end{threeparttable}
\end{table}

\subsection{Heterogeneity}
Previous research \citep[e.g.,][]{farber2015taxi} suggests that labor supply elasticities vary with experience, with less experienced workers responding less to earnings opportunities. Table \ref{tab:hetero} presents results from heterogeneity analyses based on sellers’ baseline selling activity and age groups. Baseline selling activity is measured in three ways: (1) by the number of magazines they sold during the last month (standardized with a mean of zero and a standard deviation of one), (2) by the number of magazines sold during the last year (standardized with a mean of zero and a standard deviation of one), and (3) by a dummy variable including sellers that sold more magazines than the median seller during the past three months, as described in the pre-analysis plan. As before, we use the post-double LASSO selection approach with a full set of potential control variables. We describe these measures as baseline selling activity rather than as experience, because prior sales are not a clean proxy for time spent in the occupation: they also reflect productivity, the quality of a seller's pitch and selling location, an established customer base, health, and the headroom a seller has to expand output. What the estimates establish is that, initially, more active sellers responded more strongly in levels, not that experience, as such, is the source of the heterogeneity.

\begin{table}[p]
\caption{\label{tab:hetero} Heterogeneity}
\vspace{-0.8em}
\begin{threeparttable}
\footnotesize
\renewcommand{\arraystretch}{0.85}
\setlength{\abovetopsep}{0.1ex}
\setlength{\belowbottomsep}{0.1ex}
\setlength{\aboverulesep}{0.2ex}
\setlength{\belowrulesep}{0.2ex}

\begin{subtable}[h]{\textwidth}
\caption{\label{tab:h_a} Sales last month (std)}
\resizebox{\textwidth}{!}{\begin{tabular}{l*{6}{l}}
\toprule
            &\multicolumn{1}{c}{(1)}&\multicolumn{1}{c}{(2)}&\multicolumn{1}{c}{(3)}&\multicolumn{1}{c}{(4)}&\multicolumn{1}{c}{(5)}&\multicolumn{1}{c}{(6)}\\
            &\multicolumn{1}{c}{Sales}&\multicolumn{1}{c}{ln(Sales)}&\multicolumn{1}{c}{Electronic}&\multicolumn{1}{c}{Cash}&\multicolumn{1}{c}{Hours}&\multicolumn{1}{c}{Days}\\
\midrule
Treated     &       25.32***&        0.31   &       23.45***&        0.52   &       11.89***&        2.59** \\
            &      (8.68)   &      (0.21)   &      (8.30)   &      (3.24)   &      (3.31)   &      (1.03)   \\
Sales last month&       89.64***&        0.21*  &      -36.76   &        8.77*  &       -3.20   &        0.80   \\
            &     (13.58)   &      (0.11)   &     (39.43)   &      (4.67)   &      (4.40)   &      (0.64)   \\
Treated*last month&       53.52***&        0.01   &       42.37***&       -4.08   &       15.46***&        1.06   \\
            &      (9.10)   &      (0.12)   &      (8.55)   &      (3.68)   &      (3.79)   &      (0.66)   \\
\midrule
Mean for controls&      110.68   &        3.54   &       93.57   &       17.11   &       27.11   &       10.66   \\
N           &         109   &         109   &         109   &         109   &         109   &         109   \\
\bottomrule
\end{tabular}
}
\end{subtable}
\vspace{-0.5em}

\begin{subtable}[h]{\textwidth}
\caption{\label{tab:h_b} Sales last year (std)}
\resizebox{\textwidth}{!}{\begin{tabular}{l*{6}{l}}
\toprule
            &\multicolumn{1}{c}{(1)}&\multicolumn{1}{c}{(2)}&\multicolumn{1}{c}{(3)}&\multicolumn{1}{c}{(4)}&\multicolumn{1}{c}{(5)}&\multicolumn{1}{c}{(6)}\\
            &\multicolumn{1}{c}{Sales}&\multicolumn{1}{c}{ln(Sales)}&\multicolumn{1}{c}{Electronic}&\multicolumn{1}{c}{Cash}&\multicolumn{1}{c}{Hours}&\multicolumn{1}{c}{Days}\\
\midrule
Treated     &       15.43   &        0.31   &       22.49** &        0.82   &       11.95***&        2.59** \\
            &      (9.63)   &      (0.21)   &      (9.43)   &      (3.39)   &      (3.46)   &      (1.02)   \\
Sales last 12 months&     -361.89** &        0.22** &      -44.32   &       15.23***&       -1.23   &        0.64   \\
            &    (151.99)   &      (0.11)   &     (33.66)   &      (5.57)   &      (3.55)   &      (0.59)   \\
Treated*last 12 months&       28.18** &       -0.05   &       33.24***&       -4.48   &       16.58***&        0.78   \\
            &     (11.81)   &      (0.11)   &     (12.54)   &      (4.55)   &      (4.36)   &      (0.71)   \\
\midrule
Mean for controls&      110.68   &        3.54   &       93.57   &       17.11   &       27.11   &       10.66   \\
N           &         109   &         109   &         109   &         109   &         109   &         109   \\
\bottomrule
\end{tabular}
}
\end{subtable}
\vspace{-0.5em}

\begin{subtable}[h]{\textwidth}
\caption{\label{tab:h_c} Sales last three months dummy (median 120)}
\resizebox{\textwidth}{!}{\begin{tabular}{l*{6}{l}}
\toprule
            &\multicolumn{1}{c}{(1)}&\multicolumn{1}{c}{(2)}&\multicolumn{1}{c}{(3)}&\multicolumn{1}{c}{(4)}&\multicolumn{1}{c}{(5)}&\multicolumn{1}{c}{(6)}\\
            &\multicolumn{1}{c}{Sales}&\multicolumn{1}{c}{ln(Sales)}&\multicolumn{1}{c}{Electronic}&\multicolumn{1}{c}{Cash}&\multicolumn{1}{c}{Hours}&\multicolumn{1}{c}{Days}\\
\midrule
Treated     &       -3.91   &        0.13   &       -2.94   &        0.87   &        0.65   &        1.28   \\
            &      (8.76)   &      (0.34)   &      (9.34)   &      (1.75)   &      (3.52)   &      (1.56)   \\
High seller &      -30.32** &       -0.09   &      -28.05** &        9.59   &      -14.83***&       -1.93   \\
            &     (13.58)   &      (0.40)   &     (11.76)   &      (6.34)   &      (4.48)   &      (1.77)   \\
Treated*High seller&       50.28***&        0.35   &       46.41***&       -2.64   &       22.51***&        2.63   \\
            &     (18.56)   &      (0.39)   &     (16.91)   &      (6.86)   &      (6.92)   &      (1.94)   \\
\midrule
Mean for controls&      110.68   &        3.54   &       93.57   &       17.11   &       27.11   &       10.66   \\
N           &         109   &         109   &         109   &         109   &         109   &         109   \\
\bottomrule
\end{tabular}
}
\end{subtable}
\vspace{-0.5em}

\begin{subtable}[h]{\textwidth}
\caption{\label{tab:h_d} Age (std)}
\resizebox{\textwidth}{!}{\begin{tabular}{l*{6}{l}}
\toprule
            &\multicolumn{1}{c}{(1)}&\multicolumn{1}{c}{(2)}&\multicolumn{1}{c}{(3)}&\multicolumn{1}{c}{(4)}&\multicolumn{1}{c}{(5)}&\multicolumn{1}{c}{(6)}\\
            &\multicolumn{1}{c}{Sales}&\multicolumn{1}{c}{ln(Sales)}&\multicolumn{1}{c}{Electronic}&\multicolumn{1}{c}{Cash}&\multicolumn{1}{c}{Hours}&\multicolumn{1}{c}{Days}\\
\midrule
Treated     &       23.23** &        0.30   &       20.53** &        0.97   &       11.33***&        2.55** \\
            &      (9.25)   &      (0.21)   &      (8.64)   &      (3.34)   &      (3.43)   &      (1.07)   \\
Age         &       -9.86   &       -0.29   &      -13.46   &        3.51*  &       -5.18***&       -1.02   \\
            &      (8.82)   &      (0.20)   &      (8.63)   &      (2.11)   &      (1.92)   &      (0.65)   \\
Treated*Age &        9.58   &        0.35   &       12.96   &       -4.48   &        5.79   &        0.97   \\
            &     (12.88)   &      (0.23)   &     (12.62)   &      (3.43)   &      (4.72)   &      (1.20)   \\
\midrule
Mean for controls&      110.68   &        3.54   &       93.57   &       17.11   &       27.11   &       10.66   \\
N           &         109   &         109   &         109   &         109   &         109   &         109   \\
\bottomrule
\end{tabular}
}
\end{subtable}

\vspace{0.3em}
\noindent\begin{minipage}{\textwidth}
\scriptsize
\textit{Notes:} Each panel corresponds to a different heterogeneity variable. Sales last month and Sales last 12 months are standardized (mean zero, SD one). N = 109. Robust standard errors in parentheses. \sym{*} \(p<0.10\), \sym{**} \(p<0.05\), \sym{***} \(p<0.01\). In Panel~(b), the main effect of `Sales last 12 months' is conditioned on LASSO-selected monthly lags, creating near-collinearity; this nuisance parameter does not affect the interaction terms.
\end{minipage}

\end{threeparttable}
\end{table}

Column 1 of Table \ref{tab:hetero} presents results using sales as the outcome variable, revealing strong and statistically significant heterogeneity in treatment effects. While treated individuals with average prior sales experience a positive effect of 25.32 units (p < 0.01), the interaction term between treatment and lagged sales is also positive and significant (53.52, p < 0.01), indicating that the effect of the intervention increases with individuals’ sales performance in the previous month. Specifically, the estimated treatment effect rises to approximately 79 units for those one standard deviation above the mean in prior sales. The fitted line also implies a negative effect at sufficiently low baseline sales, and it is worth being precise about where that region lies and what it contains. Because baseline sales are strongly right-skewed, the standardized measure ranges only from $-0.56$ to $7.17$, so values one standard deviation below the mean do not occur in our sample. The implied effect crosses zero at $-0.47$ standard deviations, which corresponds to having sold about 13 copies in the preceding month; 39 of the 109 sellers lie below this point, and together they account for 7.7 percent of all copies sold during the October issue. The implied effect is not distinguishable from zero anywhere in that region: at the lowest baseline sales we actually observe, it is $-4.60$ with a standard error of 8.86. Using sales over the preceding year instead (Panel~b), the crossing point is $-0.55$ standard deviations, or about 130 copies over the year, below which 25 sellers accounting for 3.8 percent of copies are found, and the implied effect at the lowest observed value is $-2.00$ (standard error 8.84). We therefore read the negative region as a linear extrapolation into the sparse lower tail of a skewed distribution rather than as evidence that the bonus reduced sales for any group. The substantive finding is that low-activity sellers did not respond, not that they responded perversely. Turning to the additional outcomes in Columns 2–6, the results are somewhat mixed. While the treatment effect on the log-transformed sales variable (Column 2) is positive but not statistically significant, the pattern of heterogeneity remains in the expected direction, although the interaction term is not statistically significant. For electronic sales (Column 3), the treatment effect is large and significant (23.45, p < 0.01), with a strong and significant interaction term (42.37, p < 0.01), mirroring the findings from Column 1 and suggesting that the overall sales effects are largely driven by increases in electronic transactions. No significant treatment effect is found for cash sales (Column 4), and the interaction term is also insignificant, implying that the intervention did not affect cash-based transactions, which is reasonable since the intervention did not target them. In terms of labor supply, the treatment significantly increases both hours worked (Column 5) and the number of working days (Column 6), with positive and significant interaction terms in Column 5 but not in Column 6, suggesting some heterogeneity in effort responses, particularly for hours worked.

Overall, irrespective of how baseline selling activity is measured (Panels a–c), the results are consistent: treatment effects are stronger for initially more active sellers than for less active ones, which aligns with previous research on experience. Initially more active sellers sold more magazines, worked more hours, and potentially more days, whereas initially less active sellers were unaffected.\footnote{Re-expressed as implied elasticities relative to the control-group mean, the heterogeneity pattern survives: the high-activity subgroup elasticity exceeds 2, while the low-activity subgroup elasticity is near zero. The mapping from levels to elasticities is mechanical here because subgroups share the control mean, so this reframing rescales rather than changes the qualitative pattern.} However, when considering age (Panel d), we observe few differences in response to the treatment, implying that there is no clear heterogeneity corresponding to the age of the sellers. A priori, it is not obvious that such differences should exist.

Distinguishing experience from a general interest in selling many papers is difficult because of likely survivorship bias: only individuals with a relative fondness for selling papers tend to do so for a long time and rely on it as a main source of income. Thus, our results are consistent with experienced sellers responding more strongly to greater earnings potential, but they may also reflect, for instance, stronger responses from those for whom the paper is a more important source of income. More generally, baseline sales bundle together several things that would each generate the same pattern: sellers who are simply more productive, those who hold a better pitch or a more lucrative selling location, those with an established base of repeat customers, those in better health, and those who had room to expand output in the first place. A seller already selling close to the limit of local demand cannot respond much to a higher commission, whatever his experience. We therefore treat the heterogeneity as informative about who responded rather than about why, and we do not read it as isolating a return to experience. Interpreting experience in the taxi studies would seem to potentially suffer from the same problem.

\subsection{Robustness}
\textit{No control variables:} As our first robustness check, we estimate the regressions from equation~\eqref{eq:regression} without control variables. As expected, given the imbalance between treated and controls found in Appendix Table \ref{tab:balance} and lower power when not including lags with predictive power, all effect estimates are smaller and imprecisely measured in models without controls for past selling behavior; see panel a in Appendix Table \ref{tab:robustness}.

\textit{Full set of controls:} Second, we estimate the regressions from equation~\eqref{eq:regression} with the full set of control variables. This controls for the imbalance between treated and waitlist control groups, but is costly given our small sample size; see panel b in Appendix Table \ref{tab:robustness}. The Hours and Days estimates are similar to those in Table \ref{tab:main}, but the estimate for Sales attenuates substantially and loses statistical significance, indicating the primary outcome's fragility to control choice.

\textit{Low-dimensional ANCOVA benchmark:} Third, to assess whether our main results are sensitive to the use of machine–learning–based covariate selection, Table~\ref{tab:sensitivity} reports transparent low-dimensional benchmarks. Column (1) presents an ANCOVA specification controlling only for the pre-treatment average of sales, yielding a positive and statistically significant treatment effect of 32.76 (p < 0.05). Column (2) shows that a parsimonious model controlling for a single one-period lag of sales (Sales$_{t-1}$) and basic demographics also delivers a positive and significant estimate of 25.21 (p < 0.05). Column (3) reproduces the main post-double LASSO estimate (23.25, p < 0.05) for comparison. The adjusted estimates are positive and statistically significant, but the estimate with the full set of controls is smaller and less precise. Covariate choice therefore matters for magnitude and precision.

\textit{Influential observations:} Fourth, we repeat the pre-registered post-double-LASSO specification after excluding one seller at a time (Appendix Figure~\ref{fig:loo}). The full-sample estimate is 23.25 (SE 9.33). All leave-one-out estimates are positive, but excluding seller 92, the largest seller in the sample, reduces the estimate to 11.75 (SE 8.05, $p=0.14$). Excluding any other seller gives estimates between 16.87 and 27.06. The positive sign is not attributable to a single seller, but the magnitude and statistical precision depend materially on the largest seller.

\textit{Permutation $p$-values:} Because $N = 109$ and the sales distribution has a long right tail (max $= 1{,}555$), asymptotic robust standard errors could overstate precision. Appendix Table~\ref{tab:permutation_pvals} reports randomization-inference $p$-values from re-randomizing treatment assignment 2{,}000 times under the actual coin-flip design, on the same parsimonious ANCOVA specification as Column~(2) of Table~\ref{tab:sensitivity}. The permutation $p$-values closely match the asymptotic ones across all six outcomes, so the inference does not appear to be driven by the long right tail.

\textit{Alternative effort measure:} Because both Hours and Days are constructed from electronic transactions, they may be mechanically affected by the bonus, which incentivizes Swish use. As complementary effort measures, we also estimate effects on sales per active day (Sales / Days) and sales per Hour, conditional on having at least one active day or one positive Hours observation. Neither ratio is independent of transaction timing, since Hours and Days are themselves built from it; what they ask is whether output rose by more than the measured work window, which is the intensity margin of interest. Both measures move only modestly under treatment (Appendix Table~\ref{tab:alt_hours}), suggesting that the increase in total sales operates primarily through the extensive margin (more days, longer spans) rather than through higher per-hour intensity. This is consistent with both standard wage-response models and the within-day measurement-error story for Hours, but it runs counter to an interpretation in which the treatment intensifies effort within a fixed work window.

\textit{Large share of electronic sales:} Following our pre-analysis plan, we estimate the effects on Hours and Days on a sample of sellers with a large share of electronic sales during the past three months (more than the median, which was 0.80). The reason for also using this sample to analyze working time is that working time measures should be more accurate for sellers who primarily sell via the electronic system. For someone selling a large fraction of their magazines for cash, working days and working time, as estimated by electronic sales, will not be accurate. As an additional check that does not rely solely on the share, we also restrict the sample to sellers above the median in absolute Swish volume during the past three months (more than 93 magazines, distinct from the unconditional median of 120 used for the heterogeneity dummy in Table~\ref{tab:hetero}, Panel~(c)). The two restrictions are imposed independently rather than in a nested fashion, so each subsample contains roughly half of the 109 experimental sellers. Estimates for Hours and Days are at least as pronounced as in the full sample, as shown in Appendix Table \ref{tab:swishshare}. Since these subsamples likely have more precise measures of working time, the results lend greater credibility to the findings in the full sample.

\textit{Within-issue dynamics:} If sellers update their effective income reference point upward in response to higher daily earnings, the treatment effect should be negative at the start of the bonus period and rise as treated sellers adapt their reference points \citep{koszegi2006model}. Appendix Table~\ref{tab:within_issue_event} reports separate ATE estimates for daily Swish copies sold within five week-long windows in the October bonus period. The treatment effect is small and insignificant in the first week (2.9, ns), jumps in the second week (7.8, $p<0.05$), and remains positive but stable in weeks three through five. The point estimates rise from the first to the second week, but the profile is not distinguishable from a flat one: a joint test of equality across the five windows gives $F(4,104) = 0.78$ ($p = 0.54$), and the first week differs from the remaining four by $-4.9$ copies (SE 5.9), clustering by seller. The first-week interval, $[-5.8, 11.6]$, includes the negative effect that reference-point updating predicts. We therefore read the within-issue profile as uninformative about the choice between a slow-emergence account and a reference-point-updating one, rather than as evidence against the latter.

\textit{Working capital and restocking:} An alternative reading of the intervention is that it operated by relaxing working-capital constraints rather than by raising the return to effort. Sellers must purchase copies up front, hold very limited cash, and wait about 45 minutes before electronic proceeds can be collected at the office (Section~2.1), so a bonus paid on electronic sales also injects liquidity into exactly this cycle. A large literature shows that households with little liquidity respond strongly to temporary income flows \citep{johnson2006household, kaplan2014model, ganong2019consumer}, which makes liquidity relief a first-order alternative interpretation rather than a technicality. The purchase register lets us test it directly: for the October and November issues, we observe each seller's individual restocking transactions, giving the number of trips to the office, the number of copies bought, and their ratio, the batch size. If the bonus relaxed a financing constraint, sellers should have bought in larger batches, making fewer trips per copy; if it raised labor supply, output and purchases rise, but the batch size, set by the cash a seller can put on the counter, should not. Because each seller holds the bonus in exactly one of the two issues, the estimates are within-seller (Appendix Table~\ref{tab:restocking}). Sellers with an active bonus made 2.31 more trips (no-bonus mean 9.93) and bought 16.02 more copies (no-bonus mean 101.12), while the batch size did not rise (point estimate $-0.18$ copies per trip against a no-bonus mean of 10.65). Conditional Poisson specifications (Appendix Table~\ref{tab:restocking_rate}) sharpen this finding: trips rose by 23.5 percent, and copies by 16.7 percent, so restocking became 9.5 percent \emph{more} frequent per copy purchased ($p = 0.005$), the opposite of what inventory financing predicts. Splitting by baseline scale (Appendix Table~\ref{tab:batch_het}), the batch-size response is $+1.56$ copies among low-baseline sellers and $-1.60$ among high-baseline sellers; neither is individually significant, but their difference is ($p = 0.025$), and the sign among the small sellers is the sign inventory financing predicts. We therefore conclude that, on average, the bonus did not change how sellers financed their inventory, which is what a labor-supply response looks like; we cannot rule out a financing response among the smallest sellers, but they account for a small share of copies sold and cannot drive the aggregate effect. A second, independent argument points the same way: decomposing the seasonally differenced sales response into purchases and returns (Appendix Table~\ref{tab:intended}, discussed below), the entire effect sits in copies purchased while returns are essentially unchanged, so treated sellers bought and sold more rather than selling down an unchanged inventory.

\textit{Total seller output:} A natural concern with the experimental setting is that treated sellers might simply have ``crowded out'' control sellers, shifting business toward themselves rather than expanding total output among the experiment participants. We address this by aggregating Swish copies sold across the 109 experimental sellers by issue cycle and comparing each cycle to the same issue one year earlier (Appendix Table~\ref{tab:total_office_sales}). Year-on-year growth among the 109 sellers was 4.2 percent for the September issue, the last cycle before any bonus, 40.3 percent for the October issue, and $-6.5$ percent for the November issue. The cycles paired across years are not of equal length, because the release calendar shifts from year to year: the October 2024 cycle ran 35 days against 28 days for its 2023 counterpart, and the November 2024 cycle 28 days against 35. Part of the October rise and part of the November fall are due to cycle length rather than output, and the year-on-year comparison is, in any case, not randomized. What it does show is that the experiment-seller totals do not contract during the October bonus cycle, which is what pure reallocation among the 109 sellers would produce. The length-controlled version of the same comparison, a seasonal difference-in-differences over a common 28-day window, is reported in Appendix Table~\ref{tab:seasonal} and is the estimate we rely on. Two further caveats are in order: the 2023 baseline excludes some non-experimental factors (price levels, seller composition), and the November cycle overlaps with the announced price increase of November 18.

The anonymized payment identifiers in the electronic sales register allow a more direct test of whether treated sellers took business from control sellers, or whether the two colluded to route sales through bonus-holders. We classify each buyer with at least two purchases from experimental sellers over the twelve pre-treatment issues by their modal (``home'') seller; there are 9{,}536 such repeat buyers, and they account for 28.6 percent of the October transactions. Appendix Table~\ref{tab:customers} tracks the share of their purchases made at October-treated sellers. Buyers whose home seller is in the waitlist group did buy more from treated sellers in October ($+6.5$ percentage points on a baseline of 6.5 percent), but the same shift is present in November ($+6.1$ points), after the bonus had switched to the waitlist group; bonus-chasing predicts a reversal, and the October--November difference for these buyers is 0.4 points (SE 1.2). The gradual, bonus-invariant drift away from home sellers in both groups is what decay of a pre-period classification looks like, not what business-stealing or collusion looks like. The scale is small: the excess of waitlist-home buyers' October purchases at treated sellers amounts to roughly 100 transactions, against an estimated treatment effect of roughly 1{,}300 copies, so even attributing all of it to displacement would account for under a tenth of the effect. That comparison covers classified repeat buyers, who make up 28.6 percent of October transactions; it is not a bound on displacement among occasional or first-time buyers, for whom no home seller can be defined.

A second concern of this general kind can be dismissed more cleanly. In many experiments on labor supply one must worry not only that treated subjects divert business from controls, but also that they shift the \emph{recording} of output across time, so that a rise in measured output during the bonus period is offset by a fall afterward and reflects re-timing rather than effort. That mechanism is largely absent here, because the product changes every month. A copy of the October issue cannot be sold in November: unsold copies must be returned to the office when the next issue is released, and the customer who buys the October issue is buying a different magazine from the one on sale a month later. A seller could not, therefore, meet a bonus target by holding inventory across the boundary or by booking the sale of an unsold copy in a later cycle. What monthly turnover does not rule out is retiming on the demand side: a customer may bring forward a periodic support purchase, buying the October issue rather than a later one, even though the two issues contain different material. Whatever the treated sellers did to sell 21 percent more copies of the October issue, it had to involve selling more copies of that issue to more customers within its own selling cycle. This does not tell us which margin of effort was used, but it does mean that the estimated effect is on real output rather than an artifact of when sales were booked; it does not, by itself, establish that none of those sales were drawn forward from later issues.

\textit{Anticipation by the waitlist group:} As flagged in Section~2.1, control sellers knew their bonus would arrive with the November issue, so the October contrast compares immediate with delayed treatment: a waitlist seller expecting a November bonus might postpone effort, which would inflate the October estimate. An external, never-informed comparison group does not exist in these data, for a mechanical reason: every Stockholm seller who entered the office during the intervention was invited. Of the 99 sellers who appear in the office's purchase register over the fourteen issue cycles without being in the experiment, only nine bought copies of the October issue. Five sellers declined to participate, and one is based outside Stockholm; of the remaining three, one is the office's own cash-sales account rather than a seller (223 copies), and the other two appear in the register for that single issue only (4 and 50 copies). We therefore probe anticipation within the experimental sample in three ways. First, we undertake a seasonal difference-in-differences comparison of each arm's change from the August-released to the September-released issue in 2024, with its own change across the same period in 2023, using electronic sales measured over a common 28-day window for all four issues (Appendix Table~\ref{tab:seasonal}). Postponement predicts a negative entry for the waitlist arm; its seasonally differenced change is $-2.65$ copies (SE 6.91), indistinguishable from zero, against $+25.54$ (SE 9.30) for the treated. The between-arm difference, an experimental estimate free of any assumption on waitlist behavior on the 2023 step, is 28.18 copies (SE 11.59, $p = 0.017$); Appendix Table~\ref{tab:seasonal_robust} reports its sensitivity, including winsorized and trimmed versions that are smaller, reflecting the right-tail concentration documented above. Second, a within-cycle triple difference asks whether waitlist sellers faded in the final week of the October cycle, immediately before their own bonus began, relative to their own behavior at the end of the twelve preceding cycles (Appendix Table~\ref{tab:triplediff}). Both arms fade, by almost the same amount, and the triple interaction is 0.125 copies per day (SE 0.288): the late-cycle fade is a feature of the October cycle common to both arms, not a waitlist-specific reduction, and the treated-waitlist gap opens early in the cycle (0.698 copies per day) and holds up late, which also speaks against treated sellers merely taking time to adapt. Third, purchases of copies from the office, which we label intended sales, are the ex-ante margin: a seller planning to hold back until November should buy fewer copies up front. The same seasonal difference-in-differences on intended sales (Appendix Table~\ref{tab:intended}) gives the waitlist arm $+11.87$ copies (SE 8.12), positive rather than negative, and release-day purchases, a single-day measure immune to differences in cycle length, likewise shows no waitlist decline ($+1.42$ copies year-on-year, SE 1.65, with the share making any release-day purchase rising from 45 to 58 percent). None of these tests recovers the missing untreated counterfactual, and the confidence intervals cannot exclude modest postponement. The tests do not provide evidence that substantial postponement by the waitlist group accounts for the October treatment effect.

\textit{Effects on the waitlist control group:} Since the control group was treated after the treatment group, we can also analyze the effects of the treatment on the control group. Appendix Table~\ref{tab:waitlist} presents results for the waitlist treatment using equation~\eqref{eq:regression} and the post-double LASSO selection approach, with previously treated individuals serving as controls and the October treatment period excluded from the candidate lags. As emphasized in Section~2.1, this comparison is contaminated by potential habit formation and learning from the October treatment of the new ``controls''. We therefore do not interpret the November estimate as a clean short-term causal effect of the bonus.

The waitlist point estimates are imprecisely measured, but they are also substantially smaller than the October effects (Appendix Table~\ref{tab:cohort_test}). For this comparison, we estimate the pre-registered specification on the two cross-sections stacked, which recovers the joint variance-covariance structure required for the equality test; the October estimates stay close to those in Table~\ref{tab:main} (sales 21.4 against 23.3, hours 8.8 against 11.0) and the November estimates to those in Table~\ref{tab:waitlist}. The two cohorts' point estimates differ by roughly a factor of four for sales (21.4 against 5.5), electronic sales (18.7 against 2.9), hours (8.8 against 2.0), and days (2.3 against 0.7), but a formal test of equality does not reject for any outcome ($p = 0.29$, $0.28$, $0.27$, and $0.31$ respectively). With 109 sellers and outcome variances of this magnitude, the design cannot statistically separate the two cohort effects; what the table establishes is that the November estimates are small and imprecise, not that they are distinguishable from the October estimates. The smaller November point estimates are consistent with the contamination concerns above, but at least one institutional feature of late November also matters here: As mentioned in Section 2.1, \textit{Situation Sthlm} announced on November 18, midway through the late-treated cohort's bonus period, a permanent commission increase from 40 to 50 SEK per copy starting with the December issue (released November 27). This increase had been discussed the year before, and the Editorial Office prefers that the price increase take effect with the release of the December issue due to higher demand before Christmas (the most recent such increase was implemented in 2021). Given the seasonality of sales and previous discussions of a price increase, some sellers may have anticipated the increase before the announcement. Once the increase was announced, the only differences from the perspective of bonus-receiving sellers are the reduced quantity demanded and the lower purchase price until the release of the December issue, so that the bonus no longer purely reflects a \textit{transitory} increase in earnings potential. As a partial probe of this channel, we re-estimate the waitlist treatment effect using only the Swish copies sold from October 30 to November 17, the first 19 days of the November bonus period, before the announcement (Appendix Table~\ref{tab:waitlist_pre18}). The pre-November-18 ATE on Swish copies is 1.4 (SE 6.1), no larger than the 2.7 (SE 9.2) estimated on Swish copies over the whole of issue 14; the announcement is therefore not the obvious driver of the muted November effect, since the muting is already present in the pre-announcement window. This is consistent with the cohort response building up gradually within the bonus period (mirroring the October within-issue pattern reported above), with intertemporal substitution of sales post-announcement in the control group, or with the contamination channels noted earlier; we cannot cleanly distinguish among the three.

\textit{Analyses including both the treatment and waitlist control group:} We also estimate two-way fixed effects (TWFE) models to assess the effect of the intervention. These estimations suffer from the same problems with using the previous intervention group as controls, but it may still be interesting to estimate an average treatment effect for both groups jointly. We estimate the average treatment effect using
\begin{equation}
    Y_{it} = \alpha_i + \lambda_t + \beta T_{it} + \varepsilon_{it},
    \label{eq:twfe}
\end{equation}
where $Y_{it}$ is the outcome, $\alpha_i$ is an individual fixed effect, $\lambda_t$ is an issue fixed effect, and $T_{it}$ equals one if individual $i$ is treated during period $t$ and zero otherwise. Concretely, the early-treated cohort has $T_{it} = 1$ in period 13 (the October issue) and zero in all other periods, while the late-treated cohort has $T_{it} = 0$ in period 13 and $T_{it} = 1$ in period 14 (the November issue). The late-treated cohort, therefore, switches from control to treated between periods 13 and 14, in tension with the standard TWFE no-anticipation assumption: knowing that a bonus is coming may already shift effort across the two periods. Table \ref{tab:did_results} reports these results, pointing in the same direction as the main results.

\begin{figure}[H]
\centering
\includegraphics[width=\textwidth]{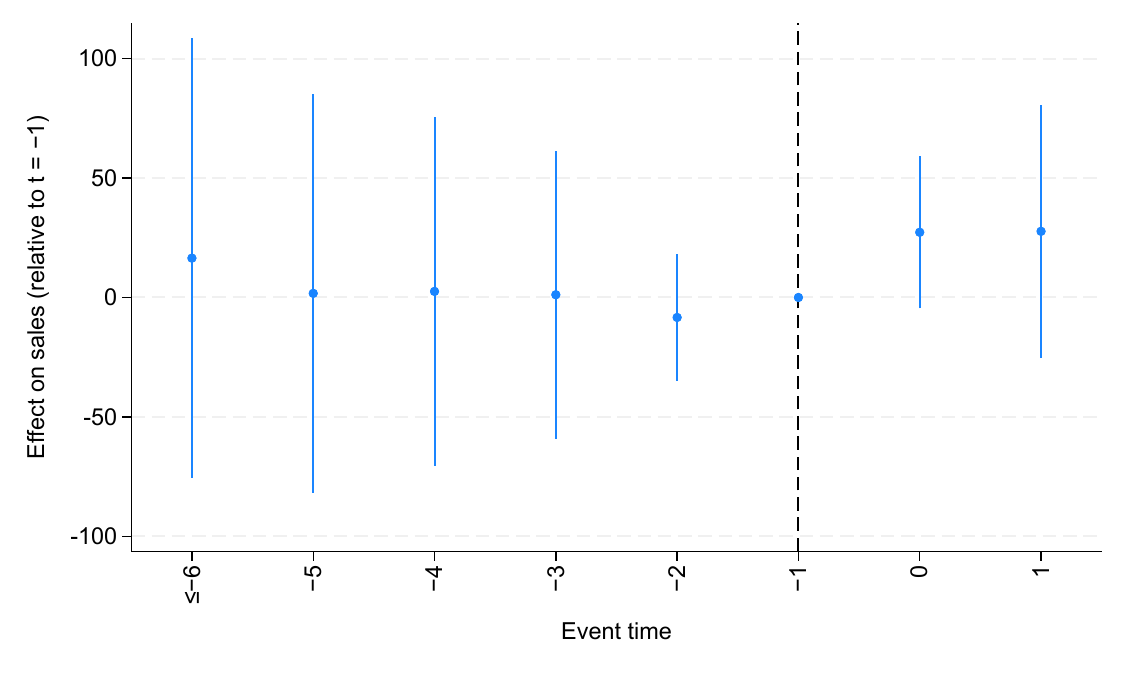}
\caption{Estimated effects on sales (event study design). Dependent variable: Sales. Sample: N = 109 sellers. Time unit: monthly issues, with $t=0$ being the treated issue. Notes: The dashed line marks the reference period ($t=-1$). Vertical bars represent 95\% confidence intervals based on standard errors clustered at the seller level.}
\label{fig:event_study}
\end{figure}

To explore potential dynamics, we also estimate a corresponding event-study specification. The pre-analysis plan envisaged estimating the regression equation~\eqref{eq:regression} separately for each period; instead, we implement an event-study version of the TWFE specification in equation~\eqref{eq:twfe}. All models include individual and period fixed effects, and standard errors are clustered at the individual level. Results are shown in Figure~\ref{fig:event_study}. The coefficients are normalized relative to the period immediately before treatment ($t=-1$), which is shown as a zero-valued reference point. Pre-treatment estimates are interpreted as evidence on pre-trends, while the estimates at $t=0$ and for post-treatment periods capture changes in sales after treatment relative to $t=-1$. The dashed vertical line marks the reference (baseline) period ($t=-1$). Confidence intervals are constructed using standard errors clustered at the individual level. The event-study results are qualitatively similar but less precisely estimated and only marginally significant. The point estimate at $t=1$ is comparable in magnitude to the effect at $t=0$; however, since the intervention was a temporary one-issue shock, this apparent persistence is likely a mechanical artifact of the standard TWFE estimator with heterogeneous cohort effects rather than genuine economic persistence \citep[cf.][]{sun2021estimating}. Because the waitlist control group shows smaller treatment effects than the initial treatment group (Table~\ref{tab:waitlist} versus Table~\ref{tab:main}), the TWFE assumption of homogeneous effects inflates the post-treatment coefficient when previously treated individuals serve as controls.

\begin{table}[H]
\centering
\begin{threeparttable}
\caption{Two-way fixed effects estimates}
\label{tab:did_results}
\small\setstretch{1}
\begin{tabular}{l*{6}{l}}
\toprule
            &\multicolumn{1}{c}{(1)}&\multicolumn{1}{c}{(2)}&\multicolumn{1}{c}{(3)}&\multicolumn{1}{c}{(4)}&\multicolumn{1}{c}{(5)}&\multicolumn{1}{c}{(6)}\\
            &\multicolumn{1}{c}{Sales}&\multicolumn{1}{c}{ln(Sales)}&\multicolumn{1}{c}{Electronic}&\multicolumn{1}{c}{Cash}&\multicolumn{1}{c}{Hours}&\multicolumn{1}{c}{Days}\\
\midrule
Treated     &       13.43** &        0.33** &       10.84*  &        2.59   &        5.41** &        1.48** \\
            &      (6.73)   &      (0.13)   &      (6.15)   &      (2.13)   &      (2.48)   &      (0.57)   \\
\midrule
Individuals &         109   &         109   &         109   &         109   &         109   &         109   \\
\bottomrule
\end{tabular}

\begin{tablenotes}[flushleft]
\small
\item \textit{Notes:} Two-way fixed effects estimates of equation~\eqref{eq:twfe}. Each column corresponds to a different outcome variable. N = 109. Standard errors clustered at the individual level in parentheses. \sym{*} \(p<0.10\), \sym{**} \(p<0.05\), \sym{***} \(p<0.01\).
\end{tablenotes}
\end{threeparttable}
\end{table}

\subsection{Survey Evidence on Motivation and Perceptions}
To complement the experimental findings, we conducted a short post-experimental survey in December 2024 (see Appendix Section~\ref{sec:surv} for the translated questionnaire). The survey was not included in the pre-analysis plan and should be interpreted as exploratory. As an incentive, respondents received two free copies of the magazine to complete the survey. Of the 109 experiment participants, 63 completed the survey (approximately 58 percent), with roughly equal numbers in the treatment (31) and control (32) groups. Respondents and non-respondents do not differ significantly on treatment status, sex, age, baseline sales, or experimental outcomes (all $p > 0.16$; Appendix Table~\ref{tab:attrition}), though respondents are modestly more active sellers in levels. The survey results should nonetheless be read as descriptive evidence for the sample of respondents.

Appendix Table~\ref{tab:surveysummary} summarizes the responses. Although 63 individuals returned the survey, the effective sample size for each item is slightly smaller due to item non-response. A majority of sellers (52 percent) report having an earnings target, and 68 percent believe that working more leads to higher sales. The questionnaire follows up with the behaviorally relevant item: whether a seller who reaches the target earlier than expected stops working or keeps selling (question~2 in Appendix Section~\ref{sec:surv}). Of the 32 respondents who report a target, only 7 (22 percent) say they would stop; the remaining 25 say they would continue selling. Only 7 of 62 respondents (11 percent) report both a target and the behavior that income targeting requires, so the headline share overstates the prevalence of income targeting in the sense that the model requires. Item non-response is not the reason for this: every respondent who reported a target answered the follow-up. The questionnaire carried no skip logic, so the item was also put to the 30 respondents who reported no target, 6 of whom nonetheless chose ``stop working''; we do not interpret those answers. Asked whether they worked more during their assigned bonus period, 16.1 percent of October-cohort respondents (5/31) said yes, against only 3.2 percent of November-cohort respondents (1 of the 31 who answered the item). The cohort gradient is qualitatively consistent with the experimental results -- the October cohort showed substantial sales increases while the November cohort did not (Section 3.4) -- but even the October-cohort rate is far below the share of treated sellers whose objective sales rose, suggesting that self-perceptions still understate the behavioral response.

Stated preferences regarding work structure and income smoothness vary across individuals. Half report working in a single stretch during the day, and 31 percent say they would prefer a lower, fixed wage over a higher, more variable one. Most respondents (66 percent) report that higher income in one month does not lead them to work less the next month, and only 25 percent report that they do not adjust their working hours based on expected earnings; the rest say they do so always or sometimes. The primary motivation for selling the magazine is financial for 63 percent of respondents.

Regarding time preferences and risk attitudes, 64 percent prefer receiving 100 SEK immediately to 150 SEK in one month, indicating high discount rates, and 72 percent prefer a certain 100 SEK to an actuarially equivalent lottery paying 200 SEK with probability one half, indicating risk aversion. These responses may help explain the general preference for immediate, stable income, though they do not appear to translate directly into income-targeting behavior in the field experiment.

We can also use the survey to look at whether the experimental treatment effect varies with stated income targeting, which is arguably the most direct survey-based test of reference dependence in the data. Appendix Table~\ref{tab:survey_target_het} estimates the treatment effect on sales and hours separately for survey respondents who report having an earnings target ($q_1 = 1$) and those who do not ($q_1 = 0$). The pattern runs against the income-targeting prediction: treatment effects are \emph{larger} for respondents who report a target (sales: 29.6, $p<0.05$; hours: 20.4, $p<0.01$) than for those who do not (sales: 7.0, ns; hours: 7.8, ns). If the income-targeting model were guiding behavior in this sample, we would expect smaller responses among targeters because the bonus would let them hit their target faster and stop working; instead, we see the opposite. Power is limited (each subgroup has roughly 30 respondents), but the point estimates are consistent with the experimental conclusion that the standard wage-response interpretation fits this population better than the income-targeting alternative. The split is based solely on the stated target. As reported above, most respondents who report a target say they would keep selling on reaching it, so $q_1$ separates sellers by whether they hold a target in mind rather than by whether they act on one; the conjunction of the two items leaves only seven respondents and cannot support a separate split.

\subsection{Observational Evidence on Reference Dependence}

As a complement to the experimental analysis, we use 12 months of pre-treatment daily Swish data to conduct observational tests of income targeting in the spirit of \citet{lalonde1986evaluating} and the taxi-driver literature \citep{camerer1997labor}. A naive fixed-effects regression of the log of daily working hours on the log of the hourly wage yields an elasticity of approximately $-1$, which, taken at face value, would suggest near-perfect income targeting. That estimate is mechanically inflated by division bias \citep{borjas1980wages}, and the two exercises that address the bias (instrumenting transitory wage variation with daily weather, and testing whether high earnings on one day reduce next-day labor supply) move the estimated elasticity toward zero or find the opposite of income targeting. The full analysis is reported in Appendix Section~\ref{app:obs}. We read it as an internal consistency check that is consistent with the experimental conclusion, not as independent identification of the wage elasticity: our setting differs from the taxi literature in important ways, the weather instruments are weak, and the next-day test has alternative explanations.

\section{Conclusion}
After randomly assigning treatment status to sellers by a coin toss, treated sellers sold more papers, the opposite of the negative response that an income-targeting view would predict. Measures of working time and days worked move in the same direction, but we treat them as suggestive because they are mechanically sensitive to the bonus's payment-mode incentive. The effect on sales appears more concentrated among sellers who were already more active at baseline. The experimental result is therefore inconsistent with the daily-income-targeting prediction in this population, while leaving the precise mechanism open: the intervention bundles a piece-rate change with payment-mode, liquidity, and salience effects.

\citet{fehrgoette} provides a particularly relevant point of comparison, since they also study labor-supply responses to a temporary increase in earnings in a randomized field experiment. In their setting (Swiss bicycle messengers paid on commission), a temporary increase in the commission rate led to a clear increase in labor supply.
Our results are consistent with the central conclusion from that experiment: when workers have substantial discretion over labor supply and face a temporary, salient improvement in earning opportunities, overall work effort increases. Our contribution is to document this in a very different labor market: one with low and volatile earnings, substantial economic constraints, and considerable discretion over labor supply, where both measurement and identification are often more challenging (as in \citet{leeson2022hobo}).

We lack a direct measure of work effort, but have a measure of hours due to the preponderance of Swish sales. According to our main point estimates, the percentage increase in hours is somewhat larger than the percentage increase in papers sold, which is qualitatively consistent with \citeauthor{fehrgoette}'s finding of reduced effort as sales per hour go down, although at the margin, effort per sale would tend to be increasing as willing buyers are exhausted. We caution that the gap is not statistically distinguishable from zero. Computed from the main post-double-LASSO point estimates in Table~\ref{tab:main}, the percentage increase in hours is 40.5\% and in sales 21.0\%, a difference of 19.5 percentage points. A bootstrap test on the unadjusted (no-controls) specification in Appendix Table~\ref{tab:main_nc}, where the corresponding gap is 17 percentage points, gives a standard error of 18.7. The unadjusted bootstrap is reported because re-bootstrapping the post-double-LASSO selection at every replication is computationally costly; the substantive conclusion (that the gap is not distinguishable from zero) is the same. Because the hours measure is imperfect and sales per hour can adjust endogenously, we do not interpret this comparison as a structural estimate of effort.

Like \citet{fehrgoette}, we rely on a randomized controlled trial rather than an observational or correlational design. Combined with sales data from before and after the intervention, this design allows us to attribute changes in observed labor supply to the bonus rather than to background trends. A further advantage of our setting is measurement. In many observational contexts, distinguishing time spent actively working from time spent taking breaks or running personal errands is difficult, which can complicate interpretation and potentially contribute to differences across studies. Our study mitigates this concern by relying on a clearly defined bonus window and objective, transaction-based measures of output, namely verified Swish sales. The RCT structure, combined with the revealed-preference nature of our data, provides a credible test of labor-supply responses to a temporary wage change in this setting.

The observational analyses are consistent with this point but should not be read as a separate identification of the wage elasticity. A naive fixed-effects regression of hours on wages, the approach pioneered by \citet{camerer1997labor}, yields an elasticity of approximately $-1$ in our data, which, taken at face value, would suggest near-perfect income targeting. This estimate is substantially inflated by division bias. When we instrument wages with daily weather, the elasticity attenuates to between $-0.2$ and $-0.4$ and loses statistical significance, paralleling \citeauthor{farber2015taxi}'s~(\citeyear{farber2015taxi}) finding for NYC taxi drivers. The weather instruments are weak in our setting, however (first-stage F-statistics in Appendix Table~\ref{tab:weather_iv} below 8), and the exclusion restriction is hard to defend for outdoor selling. A complementary test that avoids the hours-wage ratio entirely, asking whether high earnings on one day reduce next-day labor supply, finds the opposite of income targeting (good days predict \emph{more} work the next day), but is also compatible with persistence in demand, health, or seller-specific good days. We therefore treat these analyses as internal consistency checks for the experimental result rather than a decisive resolution of the income-targeting debate.

Further complicating interpretation, self-reports of motivation and behavior may also be unreliable. In our post-experimental survey, more than half of the respondents reported having income targets. Asked about effort during the bonus period, 16 percent of October-cohort respondents said they worked more, against only 3 percent of the November cohort -- a gradient in the right direction given the experimental results, but well below the objective sales increase among treated sellers. These discrepancies suggest that stated preferences may not reflect actual behavior in this sample, and caution against interpreting self-reported income targets as direct behavioral evidence that targets shape labor supply.

At least three caveats are in order. First, our intervention is embedded in a specific institutional context, and several channels beyond a pure piece-rate change could contribute to the observed effects. The Swish bonus may provide liquidity relief, allowing sellers to restock more quickly; tying the bonus to electronic sales could shift the composition of payment modes; and the bonus itself may increase the salience of earnings opportunities or perceived monitoring. While we cannot fully isolate these channels, the restocking analysis in Section~3.4 tests the liquidity channel directly and finds that, on average, sellers made more trips and bought more copies at an unchanged batch size, the pattern a labor-supply response predicts and inventory financing does not, although a financing response among the smallest sellers cannot be excluded. The absence of a significant decrease in cash sales (Table~\ref{tab:main}, Column~4) mitigates the payment-mode substitution concern, and the overall pattern (increased sales, longer hours, more working days) is consistent with a positive labor supply response regardless of the precise mechanism.

Second, because we observe only sales of the street paper, not the sellers' total labor supply across all activities, we cannot rule out the possibility that some sellers shifted effort from other income sources to paper sales. If such substitution exceeds the increase in paper sales, the overall labor-supply response could differ from what we estimate. Third, because sales per hour may decline as sellers work longer hours, the 25 percent bonus per copy need not translate into a proportional increase in the effective hourly wage; in our data, however, sales per hour and sales per active day move only modestly under treatment (Appendix Table~\ref{tab:alt_hours}), suggesting that the gap between the per-copy bonus and the effective hourly-wage change is not large. This should be kept in mind when comparing our estimates with elasticities from other settings. We offer reasons to believe our results are likely to withstand these concerns, but given an imperfect estimate of hours and the possibility that sellers derive substantial income from alternative sources, we cannot be certain.

\pagebreak
\bibliography{ref}

\pagebreak

\section*{Appendix}

\global\long\def\thetable{A\arabic{table}}
\setcounter{table}{0}
\global\long\def\theequation{A\arabic{equation}}
\setcounter{equation}{0}
\global\long\def\thefigure{A\arabic{figure}}
\setcounter{figure}{0}
\setcounter{section}{0}
\renewcommand{\thesection}{A.\arabic{section}}
\renewcommand{\theHsection}{appendix.\arabic{section}}
\renewcommand{\theHtable}{appendix.\arabic{table}}
\renewcommand{\theHfigure}{appendix.\arabic{figure}}

\section{Observational analyses (extended)}\label{app:obs}

This section reports the observational analysis summarized in Section~3.6. We do not view the exercise as a methodological correction of the taxi-driver literature (our setting differs in important ways) but as an internal consistency check that asks what the same data would say about reference dependence absent the experimental benchmark.

\subsection*{The naive hours-wage regression}

Following \citet{camerer1997labor}, we estimate fixed-effects regressions of the log of daily working hours on the log of the hourly wage, using within-seller day-to-day variation. The hourly wage is defined as $w_{id} = E_{id}/h_{id}$, where $E_{id}$ is daily earnings and $h_{id}$ is the daily Swish span (first to last transaction) for seller $i$ on day $d$. Table~\ref{tab:hours_wage} presents the results. The estimated elasticity is $-1.04$ in the full sample (Column~1) and stable across fixed-effects specifications (Columns~4 and~5), but attenuates to $-0.82$ on days with more than five sales (Column~2) and $-0.74$ on days with more than ten sales (Column~3). Taken at face value, the full-sample estimate suggests near-perfect income targeting, while the higher-volume restrictions point to a still-large but more attenuated response.

\begin{table}[H]
\centering
\begin{threeparttable}
\caption{\label{tab:hours_wage} Daily hours-wage elasticity (FE estimates)}
\small\setstretch{1}
\begin{tabular}{l*{5}{c}}
\toprule
                    &\multicolumn{1}{c}{(1)}&\multicolumn{1}{c}{(2)}&\multicolumn{1}{c}{(3)}&\multicolumn{1}{c}{(4)}&\multicolumn{1}{c}{(5)}\\
                    &\multicolumn{1}{c}{All}&\multicolumn{1}{c}{$>$5 sales}&\multicolumn{1}{c}{$>$10 sales}&\multicolumn{1}{c}{DOW FE}&\multicolumn{1}{c}{DOW+Issue FE}\\
\midrule
ln(Hourly wage)     &      -1.036\sym{***}&      -0.815\sym{***}&      -0.737\sym{***}&      -1.041\sym{***}&      -1.046\sym{***}\\
                    &     (0.022)         &     (0.042)         &     (0.072)         &     (0.020)         &     (0.019)         \\
\midrule
Observations        &       10{,}380         &        6{,}081         &        2{,}937         &       10{,}380         &       10{,}380         \\
Seller FE           &         Yes         &         Yes         &         Yes         &         Yes         &         Yes         \\
Day-of-week FE      &          No         &          No         &          No         &         Yes         &         Yes         \\
Issue FE            &          No         &          No         &          No         &          No         &         Yes         \\
\bottomrule
\end{tabular}
\begin{tablenotes}[flushleft]
\small
\item \textit{Notes:} Pre-treatment daily data (issues 1--12). Working hours are measured as the span from the first to the last Swish transaction of the day. Column~(1) includes every working day on which that span is positive; a day with a single transaction has a span of zero and therefore drops out of the log specification. Columns~(2)--(3) restrict to days with more than 5 and 10 sales. Standard errors are clustered at the seller level in parentheses: \texttt{xtreg, fe robust} reports the Arellano estimator clustered on the panel identifier, not heteroskedasticity-robust errors. \sym{*} \(p<0.10\), \sym{**} \(p<0.05\), \sym{***} \(p<0.01\).
\end{tablenotes}
\end{threeparttable}
\end{table}

However, this estimate is subject to \emph{division bias}: since the hourly wage is computed as earnings divided by hours, measurement error in hours mechanically generates a negative correlation \citep{borjas1980wages}. We address this in two ways below: first, by instrumenting transitory wage variation with daily weather; second, by testing whether yesterday's earnings predict today's labor supply (a test that avoids the hours-wage ratio entirely). Both exercises move the estimated elasticity toward zero or positive territory, paralleling \citeauthor{farber2015taxi}'s~(\citeyear{farber2015taxi}) finding that properly identified estimates for NYC taxi drivers were much smaller than the original \citet{camerer1997labor} numbers. Taken together, the observational analysis is consistent with the experimental conclusion: once division bias is addressed, the same data do not support income targeting in this population.

\subsection*{Weather as an instrument for transitory wages}

We use daily weather (temperature and precipitation) as instruments for transitory wage variation, following the logic that weather shifts buyer foot traffic and thus demand. Table~\ref{tab:weather_iv} reports the IV estimates. The point estimate using both instruments is $-0.21$ (Column~1) and the rain-only IV yields $-0.44$ (Column~3); with standard errors clustered on the seller, neither is distinguishable from zero. The attenuation from $-1.04$ to roughly $-0.2$ to $-0.4$ is consistent with \citet{farber2015taxi}, who showed that properly-instrumented estimates for NYC taxi drivers were much closer to zero than the original \citet{camerer1997labor} estimates. We note that weather may also directly affect sellers' willingness to work outdoors, complicating the exclusion restriction. The Kleibergen-Paap first-stage $F$ is low in the temperature-and-precipitation specifications (2.10 and 1.42) and stronger but still below the common rule-of-thumb threshold of 10 in the rain-only specification (7.96), so the IV estimates should be read with weak-instrument concerns in mind.

\begin{table}[H]
\centering
\begin{threeparttable}
\caption{\label{tab:weather_iv} IV estimates: Weather-instrumented hours-wage elasticity}
\small\setstretch{1}
\begin{tabular}{l*{3}{c}}
\toprule
                    &\multicolumn{1}{c}{(1)}&\multicolumn{1}{c}{(2)}&\multicolumn{1}{c}{(3)}\\
                    &\multicolumn{1}{c}{IV: Wage}&\multicolumn{1}{c}{IV: Earnings}&\multicolumn{1}{c}{IV: Rain only}\\
\midrule
ln(Hourly wage)     &      -0.207         &                     &      -0.441         \\
                    &     (0.699)         &                     &     (0.315)         \\
\addlinespace
ln(Daily earnings)  &                     &       0.238         &                     \\
                    &                     &     (0.791)         &                     \\
\midrule
Observations        &       10{,}379         &       10{,}379         &       10{,}379         \\
Kleibergen-Paap $F$ &        2.10         &        1.42         &        7.96         \\
Seller FE           &         Yes         &         Yes         &         Yes         \\
Day-of-week FE      &         Yes         &         Yes         &         Yes         \\
Instruments         &Temp, Precip         &Temp, Precip         &  Rain dummy         \\
\bottomrule
\end{tabular}
\begin{tablenotes}[flushleft]
\small
\item \textit{Notes:} IV fixed-effects regressions. Dependent variable: ln(daily working hours). Instruments are daily temperature and precipitation (Columns~1--2) or a rain indicator (Column~3). Pre-treatment data only. Estimated with \texttt{xtivreg2}. The reported first-stage statistic is the Kleibergen-Paap rk Wald $F$, computed on the same estimation sample as the second stage. Standard errors clustered at the seller level in parentheses. \sym{*} \(p<0.10\), \sym{**} \(p<0.05\), \sym{***} \(p<0.01\).
\end{tablenotes}
\end{threeparttable}
\end{table}

\subsection*{Income shocks and next-day labor supply}

We test whether high earnings on one day reduce effort the following day, a prediction of income targeting that does not involve the hours-wage ratio. We emphasize that this is a test of how a daily income shock affects next-day labor supply, not a test of intertemporal substitution at longer horizons (which our design does not identify). Since sales cluster around issue release days, we control for day-in-issue to separate release-day patterns from genuine day-to-day responses. Table~\ref{tab:intertemporal} reports the results. Across all specifications, yesterday's earnings have a \emph{positive} effect on today's labor supply. An additional 100~SEK in yesterday's earnings increases the probability of working today by 1.2 percentage points (Column~2), and a one percent increase in yesterday's earnings is associated with 0.14 percent longer hours (Column~4). These effects survive day-in-issue and issue fixed effects. Good days predict \emph{more} work the next day, not less, which is inconsistent with income targeting.

\begin{table}[H]
\centering
\begin{threeparttable}
\caption{\label{tab:intertemporal} Income shocks and next-day labor supply}
\small\setstretch{1}
\begin{tabular}{l*{4}{c}}
\toprule
                    &\multicolumn{1}{c}{(1)}&\multicolumn{1}{c}{(2)}&\multicolumn{1}{c}{(3)}&\multicolumn{1}{c}{(4)}\\
                    &\multicolumn{1}{c}{Work today}&\multicolumn{1}{c}{Work today}&\multicolumn{1}{c}{ln(Hours)}&\multicolumn{1}{c}{ln(Hours)}\\
\midrule
Yesterday's earnings (100 SEK)&       0.015\sym{***}&       0.012\sym{**} &                     &                     \\
                    &     (0.005)         &     (0.005)         &                     &                     \\
\addlinespace
ln(Yesterday's earnings) &                     &                     &       0.163\sym{***}&       0.143\sym{***}\\
                    &                     &                     &     (0.025)         &     (0.026)         \\
\midrule
Observations        &       12{,}184         &       12{,}184         &        6{,}796         &        6{,}796         \\
Seller FE           &         Yes         &         Yes         &         Yes         &         Yes         \\
Day-of-week FE      &         Yes         &         Yes         &         Yes         &         Yes         \\
Day-in-issue FE     &          No         &         Yes         &          No         &         Yes         \\
Issue FE            &          No         &         Yes         &          No         &         Yes         \\
\bottomrule
\end{tabular}
\begin{tablenotes}[flushleft]
\small
\item \textit{Notes:} Pre-treatment daily data. Columns~(1)--(2): dependent variable is whether the seller works today. Columns~(3)--(4): dependent variable is ln(hours), conditional on working. As in Table~\ref{tab:hours_wage}, standard errors are clustered at the seller level in parentheses: \texttt{xtreg, fe robust} reports the Arellano estimator clustered on the panel identifier. \sym{*} \(p<0.10\), \sym{**} \(p<0.05\), \sym{***} \(p<0.01\).
\end{tablenotes}
\end{threeparttable}
\end{table}

\clearpage

\section{Tables}\label{app:tables}

\begin{table}[H]
\centering
\begin{threeparttable}
\caption{\label{tab:summary}Summary Statistics}
\small\setstretch{1}
\begin{tabular}{lcccc}
\toprule
 & Mean & Standard Deviation & Minimum & Maximum \\
 & (1) & (2) & (3) & (4) \\
\midrule
Age & 55.83 & 10.87 & 26 & 82 \\
Woman & 0.34 & 0.48 & 0 & 1 \\
Sales & 112.85 & 191.07 & 0 & 1{,}555 \\
$\ln(\text{Sales})$ & 3.61 & 1.79 & 0 & 7.35 \\
Electronic & 96.36 & 177.08 & 0 & 1{,}467 \\
Cash & 16.50 & 25.38 & -10 & 160 \\
Hours & 30.01 & 42.68 & 0 & 286.7 \\
Days & 11.17 & 9.55 & 0 & 33 \\
\bottomrule
\end{tabular}
\begin{tablenotes}[flushleft]
\small
\item \textit{Notes:} The number of observations for all variables is 109. Sample period: October 2024 issue cycle. Units: Sales, Electronic, and Cash in copies; Hours in hours per month; Days in days per month; Age in years. ln(Sales) denotes $\ln(\text{Sales}+1)$ as defined in the pre-analysis plan.
\end{tablenotes}
\end{threeparttable}
\end{table}
\clearpage

\begin{table}[H]
\centering
\begin{threeparttable}
\caption{\label{tab:balance} Balance}
\small\setstretch{1}
\begin{tabular}{lcccc}
\toprule
 & Treatment Mean & Control Mean & Difference & P-value \\
\midrule
Age & 55.61 & 56.06 & 0.45 & 0.83 \\
Woman & 0.41 & 0.26 & -0.15 & 0.11 \\
Sales (t-1) & 78.55 & 93.23 & 14.67 & 0.62 \\
Sales (t-2) & 112.95 & 145.51 & 32.56 & 0.40 \\
Sales (t-3) & 67.73 & 92.17 & 24.44 & 0.37 \\
Sales (t-4) & 84.71 & 106.92 & 22.21 & 0.51 \\
Sales (t-5) & 61.98 & 99.77 & 37.79 & 0.18 \\
Sales (t-6) & 75.70 & 98.38 & 22.68 & 0.44 \\
Sales (t-7) & 51.73 & 94.94 & 43.21 & 0.03 \\
Sales (t-8) & 60.34 & 89.91 & 29.57 & 0.18 \\
Sales (t-9) & 135.04 & 155.87 & 20.83 & 0.69 \\
Sales (t-10) & 101.34 & 114.30 & 12.96 & 0.73 \\
Sales (t-11) & 76.04 & 97.75 & 21.72 & 0.47 \\
Sales (t-12) & 81.86 & 92.19 & 10.33 & 0.74 \\
No sales in issues 1--2 (2023) & 0.30 & 0.13 & -0.17 & 0.03 \\
\bottomrule
\end{tabular}
\begin{tablenotes}[flushleft]
\small
\item \textit{Notes:} Column 1 shows means of treated individuals ($n = 56$), Column 2 shows means of controls ($n = 53$). Column 3 reports the differences (Control minus Treatment), and Column 4 reports p-values from two-sided t-tests of mean differences. ``No sales in issues 1--2 (2023)'' is an indicator for having sold nothing in either of the two issues released one year before the intervention; it is not part of the pre-registered control set and does not enter the joint F-test. The joint F-test of balance from a regression of treatment status on the other listed variables yields F = 7.16, p < 0.01.
\end{tablenotes}
\end{threeparttable}
\end{table}
\clearpage

\begin{table}[H]
\centering
\begin{threeparttable}
\caption{\label{tab:correlations}Correlations between outcomes and lags}
\small\setstretch{1}
\begin{tabular}{l*{6}{c}}
\toprule
            & Sales & ln(Sales) & Electronic & Cash & Hours & Days \\
\midrule
(t-1)       & 0.96  & 0.74      & 0.96       & 0.69 & 0.88  & 0.80 \\
(t-2)       & 0.83  & 0.70      & 0.80       & 0.49 & 0.76  & 0.75 \\
(t-3)       & 0.89  & 0.58      & 0.90       & 0.37 & 0.77  & 0.62 \\
(t-4)       & 0.89  & 0.54      & 0.90       & 0.36 & 0.76  & 0.53 \\
(t-5)       & 0.88  & 0.51      & 0.89       & 0.34 & 0.68  & 0.55 \\
(t-6)       & 0.89  & 0.57      & 0.89       & 0.43 & 0.69  & 0.57 \\
(t-7)       & 0.58  & 0.59      & 0.57       & 0.37 & 0.46  & 0.61 \\
(t-8)       & 0.79  & 0.61      & 0.81       & 0.29 & 0.67  & 0.57 \\
(t-9)       & 0.88  & 0.52      & 0.89       & 0.39 & 0.72  & 0.58 \\
(t-10)      & 0.88  & 0.54      & 0.89       & 0.41 & 0.73  & 0.57 \\
(t-11)      & 0.88  & 0.50      & 0.90       & 0.44 & 0.71  & 0.56 \\
(t-12)      & 0.87  & 0.44      & 0.88       & 0.39 & 0.72  & 0.55 \\
\bottomrule
\end{tabular}

\begin{tablenotes}[flushleft]
\small
\item \textit{Notes:} Each column shows the correlation of the outcome variable and the lags of the outcome variable.
\end{tablenotes}

\end{threeparttable}
\end{table}
\clearpage

\begin{table}[H]
\centering
\begin{threeparttable}
\caption{\label{tab:robustness} No control variables, and full set of control variables}
\small\setstretch{1}

% --- Panel A ---
\begin{subtable}[t]{\linewidth}
\caption{\label{tab:main_nc} No control variables}
\centering
\begin{tabular}{l*{6}{l}}
\toprule
            &\multicolumn{1}{c}{(1)}&\multicolumn{1}{c}{(2)}&\multicolumn{1}{c}{(3)}&\multicolumn{1}{c}{(4)}&\multicolumn{1}{c}{(5)}&\multicolumn{1}{c}{(6)}\\
            &\multicolumn{1}{c}{Sales}&\multicolumn{1}{c}{ln(Sales)}&\multicolumn{1}{c}{Electronic}&\multicolumn{1}{c}{Cash}&\multicolumn{1}{c}{Hours}&\multicolumn{1}{c}{Days}\\
\midrule
Treated     &        4.23   &        0.14   &        5.43   &       -1.20   &        5.65   &        0.98   \\
            &     (36.43)   &      (0.34)   &     (33.74)   &      (4.92)   &      (8.12)   &      (1.84)   \\
\midrule
Mean for controls&      110.68   &        3.54   &       93.57   &       17.11   &       27.11   &       10.66   \\
N           &         109   &         109   &         109   &         109   &         109   &         109   \\
\bottomrule
\end{tabular}

\end{subtable}

\vspace{0.7em}

% --- Panel B ---
\begin{subtable}[t]{\linewidth}
\caption{\label{tab:main_fullcont} Full set of control variables}
\centering
\begin{tabular}{l*{6}{l}}
\toprule
            &\multicolumn{1}{c}{(1)}&\multicolumn{1}{c}{(2)}&\multicolumn{1}{c}{(3)}&\multicolumn{1}{c}{(4)}&\multicolumn{1}{c}{(5)}&\multicolumn{1}{c}{(6)}\\
            &\multicolumn{1}{c}{Sales}&\multicolumn{1}{c}{ln(Sales)}&\multicolumn{1}{c}{Electronic}&\multicolumn{1}{c}{Cash}&\multicolumn{1}{c}{Hours}&\multicolumn{1}{c}{Days}\\
\midrule
Treated     &       11.70   &        0.28   &       17.13*  &        1.50   &       10.13***&        3.01***\\
            &      (8.97)   &      (0.23)   &      (8.63)   &      (3.96)   &      (3.52)   &      (1.00)   \\
\midrule
Mean for controls&      110.68   &        3.54   &       93.57   &       17.11   &       27.11   &       10.66   \\
N           &         109   &         109   &         109   &         109   &         109   &         109   \\
\bottomrule
\end{tabular}

\end{subtable}

\vspace{0.6em}
\begin{tablenotes}[flushleft]
\small
\item \textit{Notes:} Each column corresponds to a different outcome variable, and each panel corresponds to a different specification. N = 109. Robust standard errors in parentheses. \sym{*} \(p<0.10\), \sym{**} \(p<0.05\), \sym{***} \(p<0.01\).
\end{tablenotes}

\end{threeparttable}
\end{table}

\begin{table}[H]
\centering
\begin{threeparttable}
\caption{\label{tab:sensitivity} Sensitivity of the treatment effect to selection of controls}
\small\setstretch{1}

\begin{tabular}{lccc}
\toprule
 & (1)  & (2)  & (3) \\
\midrule
Treated   & 32.76\sym{**} & 25.21\sym{**} & 23.25\sym{**} \\
          & (14.23) & (10.60) & (9.33)  \\
Pre-period mean sales & 1.17\sym{***} &         &         \\
          & (0.10)  &         &         \\
Sales$_{t-1}$       &         & 1.21\sym{***} & 0.89\sym{***} \\
          &         & (0.06)  & (0.09)  \\
Age       &         & -0.33   &         \\
          &         & (0.58)  &         \\
Woman     &         & -23.50\sym{**}&         \\
          &         & (10.91) &         \\
Sales$_{t-2}$ &      &         & 0.04    \\
          &         &         & (0.06)  \\
Sales$_{t-6}$ &         &         & 0.09    \\
          &         &         & (0.09)  \\
Sales$_{t-9}$ &         &         & 0.09    \\
          &         &         & (0.06)  \\
Sales$_{t-10}$ &         &         & 0.04    \\
          &         &         & (0.09)  \\
\midrule
Observations & 109 & 109 & 109 \\
\bottomrule
\end{tabular}

\begin{tablenotes}[flushleft]
\small
\item \textit{Notes:} Column (1) reports a transparent low-dimensional ANCOVA controlling for the pre-treatment average of sales, $\overline{\text{Sales}}_i = \frac{1}{12}\sum_{k=1}^{12}\text{Sales}_{i,t-k}$.
Column (2) controls for one pre-period outcome (Sales$_{t-1}$) and demographics.
Column (3) reports an OLS regression controlling for the PDS-selected controls from the post-double LASSO procedure. Robust standard errors in parentheses. The variable subscripts denote lag order, matching the convention used in Tables~\ref{tab:balance} and~\ref{tab:correlations}: Sales$_{t-1}$ is the most recent pre-treatment issue, Sales$_{t-2}$ the second most recent, and so on through Sales$_{t-12}$. \sym{*} \(p<0.10\), \sym{**} \(p<0.05\), \sym{***} \(p<0.01\).
\end{tablenotes}

\end{threeparttable}
\end{table}

\begin{figure}[H]
\centering
\includegraphics[width=0.85\textwidth]{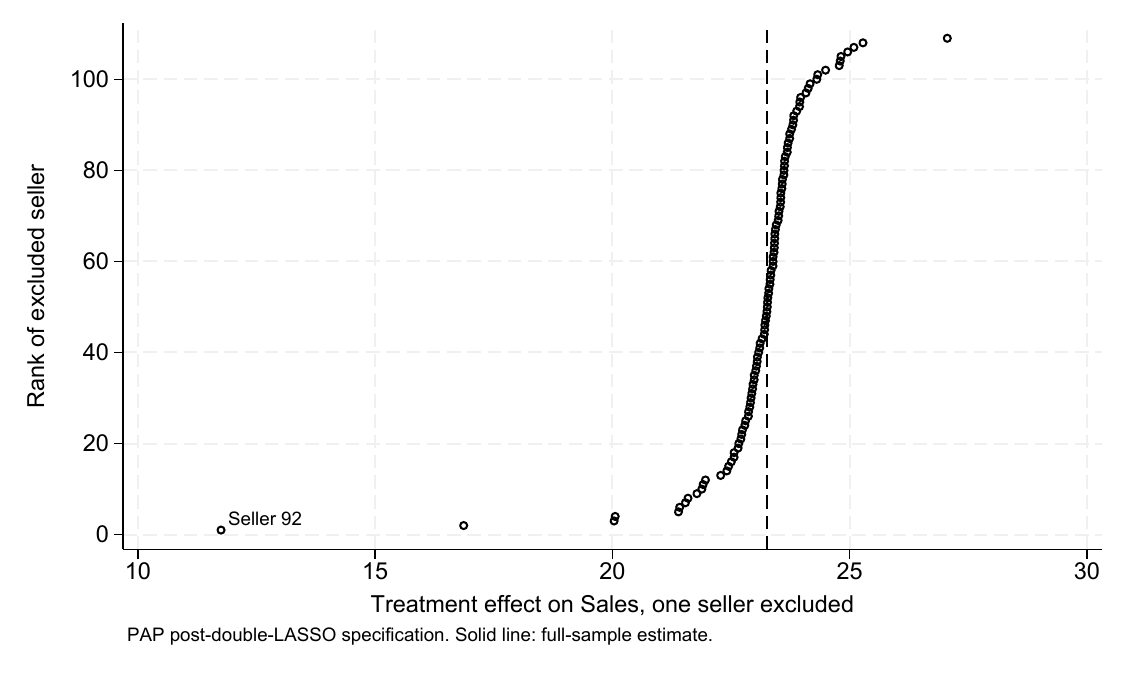}
\caption{Leave-one-seller-out estimates for the primary outcome. Notes: Each point re-estimates the pre-registered post-double-LASSO specification for Total Sales after excluding one seller. Candidate controls are age, sex, and twelve lags of Total Sales; heteroskedasticity-robust standard errors are used. The solid line marks the full-sample estimate.}
\label{fig:loo}
\end{figure}

\begin{table}[H]
\centering
\begin{threeparttable}
\caption{\label{tab:swishshare} Results for hours worked and working days}
\small\setstretch{1}

\begin{tabular}{l*{4}{c}}
\toprule
 & (1) & (2) & (3) & (4) \\
 & Hours & Days & Hours & Days \\
\midrule
Treated & 19.52\sym{**} & 4.09\sym{**} & 19.32\sym{**} & 2.82\sym{*} \\
        & (6.45)  & (1.42) & (5.90)  & (1.42) \\
\midrule
Mean for controls & 37.82 & 13.85 & 49.83 & 19.96 \\
Observations      & 54    & 54    & 54    & 54 \\
\bottomrule
\end{tabular}

\begin{tablenotes}[flushleft]
\small
\item \textit{Notes:} Columns 1--2 restrict to sellers above the median Swish share of total sales ($>$0.80) over the past three months ($N=54$). Columns 3--4 independently restrict to sellers above the median absolute Swish volume ($>$93 copies) over the same period ($N=54$). The two restrictions are not nested. Robust standard errors in parentheses. \sym{*} \(p<0.10\), \sym{**} \(p<0.05\), \sym{***} \(p<0.01\).
\end{tablenotes}

\end{threeparttable}
\end{table}
\clearpage

\begin{table}[H]
\centering
\begin{threeparttable}
\caption{\label{tab:waitlist} Results for the waitlist control, with originally treated serving as controls}
\small\setstretch{1}
\begin{tabular}{l*{6}{l}}
\toprule
            &\multicolumn{1}{c}{(1)}&\multicolumn{1}{c}{(2)}&\multicolumn{1}{c}{(3)}&\multicolumn{1}{c}{(4)}&\multicolumn{1}{c}{(5)}&\multicolumn{1}{c}{(6)}\\
            &\multicolumn{1}{c}{Sales}&\multicolumn{1}{c}{ln(Sales)}&\multicolumn{1}{c}{Electronic}&\multicolumn{1}{c}{Cash}&\multicolumn{1}{c}{Hours}&\multicolumn{1}{c}{Days}\\
\midrule
Treated     &        6.57   &        0.27   &        5.77   &        2.08   &        2.71   &        0.89   \\
            &      (9.98)   &      (0.22)   &      (9.14)   &      (1.96)   &      (3.88)   &      (0.91)   \\
\midrule
Mean for controls&       84.61   &        3.18   &       72.38   &       12.23   &       21.07   &        7.88   \\
N           &         109   &         109   &         109   &         109   &         109   &         109   \\
\bottomrule
\end{tabular}

\begin{tablenotes}[flushleft]
\small
\item \textit{Notes:} This table reports estimates of treatment effects for the main outcome, sales, and five additional outcomes that report sales or working time. Estimations were made using the post-double LASSO selection approach with a full set of potential control variables. Robust standard errors in parentheses. \sym{*} \(p<0.10\), \sym{**} \(p<0.05\), \sym{***} \(p<0.01\).
\end{tablenotes}
\end{threeparttable}
\end{table}
\clearpage

\begin{table}[H]
\centering
\begin{threeparttable}
\caption{\label{tab:cohort_test} Cohort-equality test: October vs November treatment effects}
\small\setstretch{1}
\begin{tabular}{lcccc}
\toprule
 & ATE Oct. & ATE Nov. & Difference & $p$(equal) \\
\midrule
Sales &  21.41 ( 9.89) &   5.46 (10.32) &  15.95 (15.20) & 0.294 \\
ln(Sales) &   0.35 ( 0.21) &   0.30 ( 0.22) &   0.04 ( 0.34) & 0.900 \\
Electronic &  18.74 ( 9.45) &   2.94 ( 9.69) &  15.80 (14.75) & 0.284 \\
Cash &   1.90 ( 3.80) &   3.29 ( 2.23) &  -1.39 ( 4.57) & 0.762 \\
Hours &   8.78 ( 3.72) &   2.04 ( 4.10) &   6.73 ( 6.10) & 0.269 \\
Days &   2.29 ( 1.03) &   0.66 ( 0.93) &   1.63 ( 1.60) & 0.309 \\
\bottomrule
\end{tabular}

\begin{tablenotes}[flushleft]
\small
\item \textit{Notes:} Each row is a separate estimation of equation~\eqref{eq:regression} with cohort-specific treatment indicators, using the pre-registered candidate control set (age, sex, and twelve lags of the outcome) and post-double LASSO selection, on the two cross-sections stacked ($N = 218$ seller-issues on 109 sellers). ATE Oct.\ is the treatment effect for the October cohort during issue 13, with the November cohort serving as control; ATE Nov.\ is the effect for the November cohort during issue 14, with the October cohort serving as control. An issue-14 indicator is included unpenalized. The Difference column reports ATE Oct.\ $-$ ATE Nov.\ with its standard error, and the $p$-value is from a Wald test of equality of the two ATEs. Standard errors are clustered at the seller level, in parentheses: each seller contributes an observation to both periods, so the pre-analysis plan's heteroskedasticity-robust errors are not appropriate for the stacked estimation.
\end{tablenotes}
\end{threeparttable}
\end{table}
\clearpage

\begin{table}[H]
\centering
\begin{threeparttable}
\caption{\label{tab:permutation_pvals} Asymptotic vs permutation $p$-values for main outcomes}
\small\setstretch{1}
\begin{tabular}{lcccc}
\toprule
Outcome & Coef. & Robust SE & Asymp. $p$ & Perm. $p$ \\
\midrule
Sales &  25.21 &  10.60 & 0.019 & 0.011 \\
ln(Sales) &   0.26 &   0.24 & 0.287 & 0.277 \\
Electronic &  23.13 &  10.10 & 0.024 & 0.019 \\
Cash &   1.53 &   3.43 & 0.657 & 0.667 \\
Hours &   9.18 &   3.99 & 0.023 & 0.013 \\
Days &   2.41 &   1.07 & 0.026 & 0.025 \\
\bottomrule
\end{tabular}

\begin{tablenotes}[flushleft]
\small
\item \textit{Notes:} Each row is a separate ANCOVA regression of the outcome on the treatment indicator and one pre-period lag (Sales$_{t-1}$, etc.), age, and sex, on the experimental sample ($N = 109$). The asymptotic $p$-value uses heteroscedasticity-robust standard errors; the permutation $p$-value is computed by re-randomizing treatment status 2,000 times under the actual coin-flip design and recording the share of permuted samples whose absolute estimated coefficient is at least as large as in the realized sample. Closely matching $p$-values across the two columns indicate that the asymptotic inference is not driven by the long right tail of the outcome distributions.
\end{tablenotes}
\end{threeparttable}
\end{table}
\clearpage

\begin{table}[H]
\centering
\begin{threeparttable}
\caption{\label{tab:survey_target_het} Heterogeneity by stated income target}
\small\setstretch{1}
\begin{tabular}{l*{4}{c}}
\toprule
                    &\multicolumn{1}{c}{Sales: target}&\multicolumn{1}{c}{Sales: no target}&\multicolumn{1}{c}{Hours: target}&\multicolumn{1}{c}{Hours: no target}\\
\midrule
Treated             &       29.61** &        6.98   &       20.38***&        7.81   \\
                    &     (14.06)   &     (16.00)   &      (6.45)   &      (7.69)   \\
\midrule
Mean for controls   &      105.74   &      187.83   &       26.35   &       39.53   \\
N                   &          32   &          30   &          32   &          30   \\
\bottomrule

\end{tabular}
\begin{tablenotes}[flushleft]
\small
\item \textit{Notes:} Treatment effects on sales (Columns~1--2) and hours worked (Columns~3--4) estimated separately by self-reported income target from the post-experimental survey. ``Has target'' covers respondents who answered yes to question $q_1$; ``No target'' covers those who answered no. Estimation uses the post-double LASSO selection approach with the full set of pre-period lags as candidate controls. Robust standard errors in parentheses. Sample sizes are smaller than 63 because of item non-response and absence of survey responses for non-respondents. \sym{*} \(p<0.10\), \sym{**} \(p<0.05\), \sym{***} \(p<0.01\).
\end{tablenotes}
\end{threeparttable}
\end{table}
\clearpage

\begin{table}[H]
\centering
\begin{threeparttable}
\caption{\label{tab:alt_hours} Alternative effort measures: sales per active day and per hour}
\small\setstretch{1}
\begin{tabular}{l*{2}{c}}
\toprule
                    &\multicolumn{1}{c}{Sales / active day}&\multicolumn{1}{c}{Sales / hour}\\
\midrule
Treated             &        0.48   &        1.38   \\
                    &      (0.73)   &      (1.42)   \\
\midrule
Mean for controls   &        8.59   &        8.83   \\
N                   &          85   &          73   \\
\bottomrule

\end{tabular}
\begin{tablenotes}[flushleft]
\small
\item \textit{Notes:} Column~1 reports the treatment effect on sales per active day (Sales / Days), conditional on having at least one active day; Column~2 reports the effect on sales per Hour, conditional on having a non-zero Hours observation. Estimation is OLS with one pre-period lag of the same intensity measure, age, and sex as controls, with heteroscedasticity-robust standard errors. \sym{*} \(p<0.10\), \sym{**} \(p<0.05\), \sym{***} \(p<0.01\).
\end{tablenotes}
\end{threeparttable}
\end{table}
\clearpage

\begin{table}[H]
\centering
\begin{threeparttable}
\caption{\label{tab:within_issue_event} Within-issue treatment effects on daily Swish copies sold}
\small\setstretch{1}
\begin{tabular}{l*{5}{c}}
\toprule
                    &\multicolumn{1}{c}{Days 1-7}&\multicolumn{1}{c}{Days 8-14}&\multicolumn{1}{c}{Days 15-21}&\multicolumn{1}{c}{Days 22-28}&\multicolumn{1}{c}{Days 29-35}\\
\midrule
Treated             &        2.92   &        7.84** &        3.93   &        5.47*  &        4.95** \\
                    &      (4.45)   &      (3.44)   &      (2.59)   &      (2.82)   &      (2.37)   \\
\midrule
Observations        &         105   &         105   &         105   &         105   &         105   \\
\bottomrule

\end{tabular}
\begin{tablenotes}[flushleft]
\small
\item \textit{Notes:} Each column reports a separate seller-level OLS regression of weekly Swish copies sold on the treatment indicator, controlling for one pre-period lag of monthly sales (Sales$_{t-1}$), age, and sex. The dependent variable is the sum of daily Swish copies within the indicated week-long window of the October bonus period (the issue cycle has approximately 35 days). Robust standard errors in parentheses. \sym{*} \(p<0.10\), \sym{**} \(p<0.05\), \sym{***} \(p<0.01\).
\end{tablenotes}
\end{threeparttable}
\end{table}
\clearpage

\begin{table}[H]
\centering
\begin{threeparttable}
\caption{\label{tab:total_office_sales} Total Swish copies sold by the 109 experimental sellers, by issue cycle, year-on-year}
\small\setstretch{1}
\begin{tabular}{lccccc}
\toprule
 & \multicolumn{2}{c}{2023} & \multicolumn{2}{c}{2024} & \\
\cmidrule(lr){2-3}\cmidrule(lr){4-5}
Issue & Days & Copies & Days & Copies & YoY (\%) \\
\midrule
September issue &  28 &     7,517 &  28 &     7,834 &   4.2 \\
October issue &  28 &     7,485 &  35 &    10,503 &  40.3 \\
November issue &  35 &     9,360 &  28 &     8,750 &  -6.5 \\
December issue &  28 &    12,717 &  30 &    14,514 &  14.1 \\
\bottomrule
\end{tabular}

\begin{tablenotes}[flushleft]
\small
\item \textit{Notes:} Total Swish copies sold by the 109 experiment sellers per issue cycle, comparing each 2024 issue to the same issue of 2023. An issue cycle runs from the release of one issue to the release of the next, the unit used throughout the paper, and an issue is named for the month it covers rather than the month it is released in: the October 2024 issue was released on September 25 and sold until the November issue appeared on October 30. There are eleven releases a year, since the June issue is a double issue, so the 2023 counterpart of issue $k$ is issue $k-11$. The October and November rows are the October-treated and November-treated cohorts' bonus cycles. Paired cycles differ in length, as the Days columns show, and the totals are not adjusted for that difference; Appendix Table~\ref{tab:seasonal} reports the corresponding comparison over a common 28-day window. Source: \textit{Situation Sthlm} Swish ledger.
\end{tablenotes}
\end{threeparttable}
\end{table}
\clearpage

\begin{table}[H]
\centering
\begin{threeparttable}
\caption{\label{tab:waitlist_pre18} Waitlist treatment effect on Swish sales before the November 18 announcement}
\small\setstretch{1}
\begin{tabular}{l*{3}{c}}
\toprule
                    &\multicolumn{1}{c}{Pre-Nov 18}&\multicolumn{1}{c}{Pre-Nov 18}&\multicolumn{1}{c}{Full issue 14}\\
\midrule
Treated             &        1.43   &        1.60   &        2.71   \\
                    &      (6.05)   &      (7.76)   &      (9.18)   \\
\midrule
Estimator           &   PDS-Lasso   &      ANCOVA   &   PDS-Lasso   \\
Mean for controls   &       55.09   &       55.09   &       72.38   \\
N                   &         109   &         109   &         109   \\
\bottomrule

\end{tabular}
\begin{tablenotes}[flushleft]
\small
\item \textit{Notes:} The dependent variable in Columns~1 and 2 is the total number of Swish copies sold by each seller over October 30 -- November 17, 2024 (the first 19 days of the November issue period, before the November 18 price-increase announcement). Column~3 reports the same specification with Swish copies sold over the whole of issue 14 as a benchmark, so that all three columns measure the same outcome over different windows. ``Treated'' here is the November-cohort treatment indicator: the November-treated cohort serves as the treatment group and the October-treated cohort as control. Estimation in Columns~1 and 3 uses post-double LASSO with twelve monthly lags (issues 1--12) plus age and sex as candidate controls; Column~2 reports a parsimonious ANCOVA controlling for sales$_{t-1}$, age, and sex. Robust standard errors in parentheses. \sym{*} \(p<0.10\), \sym{**} \(p<0.05\), \sym{***} \(p<0.01\).
\end{tablenotes}
\end{threeparttable}
\end{table}
\clearpage

\begin{table}[H]
\centering
\begin{threeparttable}
\caption{\label{tab:surveysummary} Descriptive statistics from survey}
\small\setstretch{1}
\begin{tabular}{l*{2}{l}}
\toprule
                    &\multicolumn{2}{c}{}     \\
                    &        Mean&       Count\\
\midrule
Having an earnings target&       0.516&          62\\
Would stop working if target reached early (has target)&       0.219&          32\\
Sell more if working more&       0.678&          59\\
Work without interruption&       0.500&          60\\
Prefer lower fixed wage&       0.309&          55\\
Would not work less next month if earned more this month&       0.656&          61\\
Do not plan working hours depending on expected earnings&       0.246&          57\\
Work mainly to earn money&       0.627&          59\\
Worked more due to intervention&       0.097&          62\\
Worked less due to the intervention&       0.048&          62\\
No change in working hours due to the intervention&       0.855&          62\\
Want 100 now instead of 150 in one month&       0.638&          58\\
Are risk averse     &       0.719&          57\\
\bottomrule
\end{tabular}

\begin{tablenotes}[flushleft]
\small
\item \textit{Notes:} See Appendix Section \ref{sec:surv} for the survey questions.
\end{tablenotes}
\end{threeparttable}
\end{table}
\clearpage

\begin{table}[H]
\centering
\begin{threeparttable}
\caption{\label{tab:restocking} Inventory purchases: within-seller estimates}
\small\setstretch{1}
\begin{tabular}{lcccc}
\toprule
 & (1) & (2) & (3) & (4) \\
 & Trips & Copies & Copies per trip & Trips per 100 copies \\
\midrule
Bonus active & 2.31\sym{***} & 16.02\sym{**} & -0.18 & 0.91 \\
 & (0.81) & (6.95) & (0.72) & (1.44) \\
\midrule
Mean, no-bonus months & 9.93 & 101.12 & 10.65 & 21.22 \\
N (seller-issues) & 218 & 218 & 196 & 196 \\
Sellers identifying & 109 & 109 & 91 & 91 \\
\bottomrule
\end{tabular}

\begin{tablenotes}[flushleft]
\small
\item \textit{Notes:} Each seller is observed in the October and November issues, and receives the bonus in exactly one of them. All columns include seller and issue fixed effects; standard errors clustered on the seller in parentheses. Trips is the number of separate purchase transactions at the office. Copies per trip and Trips per 100 copies are undefined for a seller-issue with no purchases, which is why columns (3) and (4) use fewer observations; the final row reports the number of sellers who contribute to both observations and therefore identify the within-seller estimate. \sym{*} \(p<0.10\), \sym{**} \(p<0.05\), \sym{***} \(p<0.01\).
\end{tablenotes}
\end{threeparttable}
\end{table}
\clearpage

\begin{table}[H]
\centering
\begin{threeparttable}
\caption{\label{tab:restocking_rate} Inventory purchases: multiplicative specifications}
\small\setstretch{1}
\begin{tabular}{lcccc}
\toprule
 & (1) & (2) & (3) & (4) \\
 & Trips & Copies & Trips per copy & Copies per trip \\
 & Poisson & Poisson & Poisson, offset & OLS, zeros kept \\
\midrule
Bonus active & 0.211\sym{***} & 0.154\sym{**} & 0.090\sym{***} & 0.444 \\
 & (0.067) & (0.060) & (0.032) & (0.683) \\
\midrule
Implied \% change & 23.5 & 16.7 & 9.5 & --- \\
N (seller-issues) & 210 & 210 & 182 & 218 \\
\bottomrule
\end{tabular}

\begin{tablenotes}[flushleft]
\small
\item \textit{Notes:} Columns (1)--(3) are conditional (fixed-effects) Poisson specifications with issue fixed effects and standard errors clustered on the seller; coefficients are log points and the implied percentage change is $100(e^{\beta}-1)$. Column (3) includes the log of copies purchased as an offset, so the coefficient is the effect on trips per copy. Column (4) is the linear specification of Table~\ref{tab:restocking} column (3) with batch size set to zero for seller-issues without purchases, so that no observation is dropped. Poisson specifications drop sellers with no variation in the outcome. \sym{*} \(p<0.10\), \sym{**} \(p<0.05\), \sym{***} \(p<0.01\).
\end{tablenotes}
\end{threeparttable}
\end{table}
\clearpage

\begin{table}[H]
\centering
\begin{threeparttable}
\caption{\label{tab:batch_het} Batch size, by baseline selling scale}
\small\setstretch{1}
\begin{tabular}{lccc}
\toprule
 & (1) & (2) & (3) \\
 & Low baseline & High baseline & Difference \\
\midrule
Bonus active & 1.56 & -1.60 & 3.15\sym{**} \\
 & (1.00) & (1.00) & (1.40) \\
\midrule
Mean, no-bonus months & 5.11 & 15.53 & \\
N (seller-issues) & 91 & 105 & 196 \\
\bottomrule
\end{tabular}

\begin{tablenotes}[flushleft]
\small
\item \textit{Notes:} Outcome is copies per restocking trip. Specification as in Table~\ref{tab:restocking} column (3). Low and high baseline are defined by the median of total sales over the three pre-treatment issues. Column (3) reports the interaction between the bonus indicator and low baseline, estimated on the pooled sample. \sym{*} \(p<0.10\), \sym{**} \(p<0.05\), \sym{***} \(p<0.01\).
\end{tablenotes}
\end{threeparttable}
\end{table}
\clearpage

\begin{table}[H]
\centering
\begin{threeparttable}
\caption{\label{tab:intended} Seasonal difference-in-differences: intended sales, sales, and returns}
\small\setstretch{1}
\begin{tabular}{lccc}
\toprule
 & (1) & (2) & (3) \\
 & Intended sales & Sales & Copies returned \\
\midrule
Waitlist arm & 11.87 & 11.89 & 0.57 \\
 & (8.12) & (8.06) & (1.12) \\
\addlinespace
Treated arm & 45.29\sym{***} & 42.18\sym{***} & 2.82 \\
 & (12.73) & (12.03) & (1.88) \\
\addlinespace
Difference (treated $-$ waitlist) & 33.42\sym{**} & 30.29\sym{**} & 2.26 \\
 & (15.10) & (14.48) & (2.19) \\
\midrule
Sellers (waitlist / treated) & \multicolumn{3}{c}{53 / 56} \\
\bottomrule
\end{tabular}

\begin{tablenotes}[flushleft]
\small
\item \textit{Notes:} Each cell is a seasonal difference-in-differences at the seller level: the change from the August-released to the September-released issue in 2024, minus the same change in 2023. Intended sales are copies purchased from the office; Sales are copies sold; Copies returned enter such that Sales $\approx$ Intended sales $-$ Copies returned. The first two rows report each arm's mean with the standard error of the mean in parentheses; the third row is the coefficient on the treatment indicator from a regression of the seller-level difference-in-differences on treatment, with robust standard errors. Postponement by the waitlist arm predicts a negative entry in the first row of column (1). \sym{*} \(p<0.10\), \sym{**} \(p<0.05\), \sym{***} \(p<0.01\).
\end{tablenotes}
\end{threeparttable}
\end{table}
\clearpage

\begin{table}[H]
\centering
\begin{threeparttable}
\caption{\label{tab:seasonal} Seasonal difference-in-differences: electronic sales}
\small\setstretch{1}
\begin{tabular}{lcccc}
\toprule
 & 2023 step & 2024 step & Difference (DiD) & $p$ \\
\midrule
Waitlist (control) & 5.86 & 3.22 & -2.65 (6.91) & 0.703 \\
Treated in October & -6.13 & 19.41 & 25.54 (9.30) & 0.008 \\
\midrule
Treated $-$ waitlist & & & 28.18 (11.59) & 0.017 \\
\bottomrule
\end{tabular}

\begin{tablenotes}[flushleft]
\small
\item \textit{Notes:} Electronic copies sold, each issue measured over its first 28 days (the October issue had a 35-day cycle, so a common window is imposed on all four issues). The 2023 step is the change from the August-released to the September-released issue of 2023; the 2024 step is the same change in 2024. The difference-in-differences column is the 2024 step minus the 2023 step, with the standard error of the seller-level mean in parentheses. The final row is the difference between arms, from a regression of the seller-level difference-in-differences on the treatment indicator with robust standard errors. 51 waitlist and 54 treated sellers; four sellers, two in each arm, have no electronic transactions in the daily file.
\end{tablenotes}
\end{threeparttable}
\end{table}
\clearpage

\begin{table}[H]
\centering
\begin{threeparttable}
\caption{\label{tab:seasonal_robust} Seasonal difference-in-differences: sensitivity}
\small\setstretch{1}
\begin{tabular}{lccc}
\toprule
 & Estimate & $p$ & N \\
\midrule
Matched 28-day windows, electronic copies & 28.18 (11.59) & 0.017 & 105 \\
\quad restricted to sellers active in 2023 & 37.82 (14.16) & 0.009 & 82 \\
\quad winsorised at the 95th percentile & 16.85 (9.82) & 0.089 & 105 \\
\quad dropping the largest seller & 24.33 (11.06) & 0.030 & 104 \\
\midrule
Total copies, unmatched windows & 30.29 (14.48) & 0.039 & 109 \\
\quad dropping the largest seller & 23.44 (12.90) & 0.072 & 108 \\
\quad winsorised at the 95th percentile & 15.92 (11.50) & 0.169 & 109 \\
\quad trimming the top and bottom 5\% & 12.61 (9.09) & 0.168 & 99 \\
\quad restricted to sellers active in 2023 & 43.83 (18.73) & 0.022 & 85 \\
\quad in logs & 0.334 (0.323) & 0.304 & 109 \\
\bottomrule
\end{tabular}

\begin{tablenotes}[flushleft]
\small
\item \textit{Notes:} Each row reports the coefficient on the treatment indicator from a regression of the seller-level seasonal difference-in-differences on treatment, with robust standard errors. The upper panel measures all four issues over a common 28-day window using electronic copies; the lower panel uses total copies over each issue's full cycle, so that the October issue spans 35 days and its 2023 counterpart 28. ``Active in 2023'' restricts to sellers with positive sales in either of the two 2023 issues. ``In logs'' uses $\ln(1+\text{Sales})$ throughout.
\end{tablenotes}
\end{threeparttable}
\end{table}
\clearpage

\begin{table}[H]
\centering
\begin{threeparttable}
\caption{\label{tab:triplediff} Within-cycle timing of effort}
\small\setstretch{1}
\begin{tabular}{lccc}
\toprule
 & Estimate & Std. err. & $p$ \\
\midrule
October cycle $\times$ late (waitlist group) & -0.592 & (0.187) & 0.002 \\
October cycle $\times$ late (treated group) & -0.467 & (0.218) & 0.035 \\
\quad triple interaction (difference) & 0.125 & (0.288) & 0.665 \\
\midrule
October cycle $\times$ Treated & 0.698 & (0.421) & 0.100 \\
\midrule
Waitlist mean, late October cycle & 1.854 & & \\
Treated mean, late October cycle & 2.103 & & \\
N (seller-days) & 44,835 & & \\
\bottomrule
\end{tabular}

\begin{tablenotes}[flushleft]
\small
\item \textit{Notes:} Outcome is electronic copies sold per seller-day over issues 1--13. Late is defined as the final seven days of each issue's own cycle. The specification includes seller fixed effects and the full set of interactions between the October-cycle, late and treated indicators; standard errors clustered on the seller. The first two rows report each arm's change in late-cycle daily sales in the October cycle relative to its own average over the twelve preceding cycles; the third row is their difference. 105 sellers (51 waitlist, 54 treated).
\end{tablenotes}
\end{threeparttable}
\end{table}
\clearpage

\begin{table}[H]
\centering
\begin{threeparttable}
\caption{\label{tab:customers} Repeat customers' purchases across treatment arms}
\small\setstretch{1}
\begin{tabular}{lcc}
\toprule
 & Home seller & Home seller \\
 & treated in October & waitlist \\
\midrule
\multicolumn{3}{l}{\emph{Share of purchases made at October-treated sellers}} \\
\quad issues 1--12 (classification window) & 0.924 & 0.065 \\
\quad October issue (bonus on treated) & 0.906 & 0.130 \\
\quad November issue (bonus on waitlist) & 0.875 & 0.126 \\
\midrule
October change, relative to issues 1--12 & -0.018\sym{**} & 0.065\sym{***} \\
 & (0.008) & (0.008) \\
November change, relative to issues 1--12 & -0.050\sym{***} & 0.061\sym{***} \\
 & (0.009) & (0.009) \\
October versus November & 0.032\sym{***} & 0.004 \\
 & (0.011) & (0.012) \\
\midrule
Buyers & 3,904 & 5,632 \\
Transactions & 20,153 & 25,362 \\
\bottomrule
\end{tabular}

\begin{tablenotes}[flushleft]
\small
\item \textit{Notes:} Repeat customers are buyers with at least two electronic purchases from experimental sellers over the twelve pre-treatment issues and a unique modal (home) seller; the buyer identity is an anonymized payment identifier, replaced at import, and no identifying information is retained in any analysis file. Cells in the upper panel are shares of the group's purchases made at October-treated sellers. Changes are estimated from a linear regression of an at-treated-seller indicator on home-arm and period interactions, with standard errors clustered on the buyer in parentheses. \sym{*} \(p<0.10\), \sym{**} \(p<0.05\), \sym{***} \(p<0.01\).
\end{tablenotes}
\end{threeparttable}
\end{table}
\clearpage

\begin{table}[H]
\centering
\begin{threeparttable}
\caption{\label{tab:attrition} Survey respondents and non-respondents}
\small\setstretch{1}
\begin{tabular}{lcccc}
\toprule
 & Respondents & Non-respondents & Difference & $p$ \\
\midrule
Treated in October & 0.49 & 0.54 & -0.05 & 0.600 \\
Woman & 0.29 & 0.41 & -0.13 & 0.169 \\
Age & 55.43 & 56.37 & -0.94 & 0.658 \\
Mean sales, issues 1--12 & 102.07 & 83.42 & 18.65 & 0.530 \\
Sales, issues 10--12 & 329.19 & 246.00 & 83.19 & 0.368 \\
Sales, October issue & 114.94 & 110.00 & 4.94 & 0.895 \\
Sales, November issue & 110.78 & 74.85 & 35.93 & 0.261 \\
\midrule
Sellers & 63 & 46 & & \\
\bottomrule
\end{tabular}

\begin{tablenotes}[flushleft]
\small
\item \textit{Notes:} Means by survey-response status; $p$-values from two-sided $t$-tests. Sales variables are copies sold. 63 respondents, 46 non-respondents. \sym{*} \(p<0.10\), \sym{**} \(p<0.05\), \sym{***} \(p<0.01\).
\end{tablenotes}
\end{threeparttable}
\end{table}
\clearpage

% Första PDF-sidan: blir A.2 Questionnaire


\section*{Supplementary material}

## A.2 Questionnaire

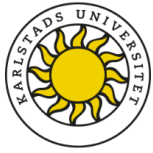

**Survey for Sellers of Situation Sthlm**

Thank you for participating in the "Swish Campaign," which was actually a study conducted by researchers at Karlstad University and funded by the Jan Wallander and Tom Hedelius Foundation and the Tore Browaldh Foundation. We would like to ask you to answer a short survey about how the Swish study affected you. As a thank-you for completing the survey, you will receive two magazines as compensation. All responses are confidential. Neither your sales during the Swish campaign nor your answers to the survey can be traced back to you as an individual.

---

**1. Do you have a specific daily or weekly income goal when selling the magazine?**
☐ Yes
☐ No

**2. If you reach your income goal earlier than expected, are you more likely to stop working or continue selling?**
☐ Stop working
☐ Continue selling

**3. Do you generally feel that you sell more magazines if you work longer?**
☐ Yes
☐ No

**4. When I sell magazines during a day (without the Swish bonus), I usually...**
☐ ...work in one stretch, maybe with a few short breaks
☐ ...work in several short shifts
☐ ...sometimes work one way, sometimes the other

**5. Did the above change during the period you received a bonus for Swish sales?**
☐ Yes, I worked more in short shifts
☐ Yes, I worked longer stretches
☐ No

**6. Would you prefer to earn a fixed amount every day or an amount that sometimes is low and sometimes high, but on average over a month is a bit higher than the fixed amount?**
☐ Fixed amount
☐ Varying amount

**7. If you earned more money during a particular month, would you be inclined to work less the following month?**
☐ Yes

□ No
□ Don't know

**8. Do you plan your working hours based on how much you think you'll earn (e.g., more when there's a new issue, less later; more when lots of people are out, less at other times)?**
□ Yes
□ No
□ Sometimes

**9. What motivates you the most to sell magazines for Situation Sthlm? Rank these from 1 (most motivating) to 4 (least motivating):**
□ Earning money
□ Meeting people
□ Personal satisfaction
□ Having something to do during the day

**10. Did you spend more or less time selling Situation Sthlm during the month you received extra money per magazine sold via Swish?**
□ I worked more
□ I worked less
□ I worked about the same as usual

**11. If you could choose, would you prefer to get 100 SEK now or 150 SEK in one month?**
□ 100 SEK now
□ 150 SEK in a month

**12. What would you choose: to receive 100 SEK for sure or a 50% chance to get 200 SEK (and 0 SEK otherwise)?**
□ 100 SEK guaranteed
□ 50% chance to get 200 SEK

# A.3 Pre-analysis plan

Pre-analysis plan for Reference Dependent or Independent Labour Supply among Street Paper Sellers

Mats Ekman, Niklas Jakobsson, and Andreas Kotsadam

September 16, 2024

**Abstract**

We investigate whether labour suppliers behave as though they have reference-dependent preferences. The present pre-analysis plan discusses how the variables are coded, how we will deal with missing values, and the specification of the estimation equations. We also conduct a power analysis.

1

## Introduction

Dynamic labour supply models suggest that work hours should increase in response to temporary wage increases. This occurs because workers shift labour and leisure over time, choosing to work more when wages are higher and to enjoy more leisure when the opportunity cost—the wage they forego—is lower. However, a permanent wage increase will trigger an offsetting income effect, as the rise in lifetime wealth allows workers to afford more leisure (Farber, 2015). We investigate whether labour suppliers behave according to the dynamic labour supply models or if they have reference-dependent preferences to attempt to reach specific income targets, which aligns with the predictions of prospect theory (Kahneman and Tversky, 1979). Our participants are sellers of the Swedish street paper *Situation Stockholm*. Our main focus is how the workers respond to transitory income shocks.

## The intervention

Our group of participants is sellers of the Swedish street paper *Situation Stockholm* (henceforth "*Situation Sthlm*" or "the paper"). Sellers buy copies from the paper's office for 40 SEK each and sell them for 80 SEK. The paper is published monthly, except for one summer month. Sellers who have unsold copies when a new issue comes out may return the copies to the paper's office and get 40 SEK per copy back.

In an intervention directed at sellers of *Situation Sthlm*, we reward sellers with ten SEK extra per copy they sell. To eliminate the possibility of cheating (e.g., sales between sellers where the issues can later be sold again), only sales made through an electronic payment method known as Swish qualify for the reward. Swish sales already comprise the vast majority of all sales, and enable us to verify that the buyers are not other sellers, as well as to exclude buyers of multiple papers of the same issue. Sellers know these rules and should expect their sales behaviour to be monitored. Participating sellers also sign a contract that promises not to cheat. Participants do not know that they are in a study, having been told that the editorial office of the paper has simply received a donation.

A coin flip randomizes sellers. In concrete terms, sellers looking to buy copies of *Situation Sthlm* are asked if they would like to flip a coin to determine if they will receive a ten-crown reward per copy sold. through Swish. If the coin comes up heads, he or she gets the reward. Otherwise, he or she gets the reward the next issue.

## Data

We use data from the editorial office of *Situation Sthlm* on all sales by individual sellers. For electronic sales, we have access to real-time sales data; for payments in cash, we will know that the issue was sold in the period between when the seller picked it up and when s/he went back to potentially return

unsold magazines (a period that may cover several days or weeks). In addition to the sales data, we also have access to the age and gender of the sellers.

## Coding of variables

In this section, we present the variables that we will use in the analysis. We start with the outcome variables and continue with the covariates and the variables used to study heterogeneous effects.

### Main outcome

Total number of sold papers of the September 25 issue: Sales can be made in cash or through the electronic system Swish. On average, about 80 percent of sales are made via Swish. Our primary outcome measure is the total number of sold papers of the September 25 issue, including cash and Swish sales.

### Other exploratory outcomes

Log of Total number of sold papers of the September 25 issue

Total number of sold papers of the September 25 issue via the electronic Swish system

Total number of sold papers of the September 25 issue via cash

Hours worked selling the September 25 issue: We can approximate the daily number of hours worked by our measure of electronic sales since we know the exact time of day for electronic sales. This measure of working hours is not an exact measure of working time, and the intervention might affect the share of sales via the electronic system. For sellers with a high share of electronic sales our measure will be better and we will consider this outcome variable for the full sample as well as for a restricted sample of sellers with a large share of electronic sales during the previous three months. We will also restrict the hours-worked analysis to days with at least two electronic sales.

Number of days worked selling the September 25 issue: We can approximate the number of days worked via electronic sales data, similar to how we estimate hours worked.

### Control variables

Our control variables are administrative data given us by the editorial office of the paper and are all measured before the intervention starts.

12 lagged issues of past number of sold papers.

Female: Equal to one if the seller is a woman.

Age: Continuous variable of age.

## Variables to measure heterogeneity

Total number of sold papers during the last 12 months

Total number of sold papers during the last month

High seller: A dummy equal to one for sellers who have sold more than the median number of the previous three issues of the paper.

Age

## Estimation strategy

We first regress *Total sales of the September 25 issue* on treatment status; i.e. a dummy variable equal to one if the individual was randomly assigned the lower price, and a set of controls, $\boldsymbol{X}_i$:

(1)     $Y_i = \alpha + \beta_1 T_i + \beta_2 \boldsymbol{X}_i + \varepsilon_i$

The vector $\mathbf{X}$ comprises the covariates listed above (Lags of the dependent variable, Female, and Age). The regression is estimated with OLS and robust standard errors.

To test for balance, we will regress our main treatment variable on the control variables described above both individually and together. We will judge whether the randomization worked by conducting an F-test of whether the control variables jointly predict treatment status.

We will present results from estimations without control variables. However, our main estimation will be one where optimal controls are chosen from the total list of controls using a post-double LASSO selection approach by Belloni, Chernozhukov, and Hansen (2014). The LASSO selection approach selects those variables that are correlated with both treatment and the outcomes, which may improve precision in the estimates (especially including the lagged values of sales), and it also helps to correct for imbalances across groups. We will also present the results using an event plot where we estimate equation (1) for each period. We will then test for balance for each individual lag and for the lags jointly.

If we have missing values on explanatory variables, we will code the variables as zero and include dummy variables controlling for missing status so that we do not lose observations. To make the models fully saturated, we partition the covariate space and add control variables as indicator variables rather than using their multi-valued codes (Athey and Imbens, 2017). If cells are too small, with less than 5 percent of the observations, adjacent cells are combined. When using interaction terms and in tests of balance, we will retain the continuous coding of the variables.

## Exploratory analyses

### Exploratory heterogeneity analyses

We will investigate whether sellers differ in their labour supply behaviour. Are some sellers reference-dependent and others optimizers? Following findings from previous research (Farber, 2015), there is evidence of more reference dependence among the inexperienced and less among the experienced. We will explore this issue by comparing experienced sellers to inexperienced ones using the variables described in "Variables to measure heterogeneity".

### Exploratory analysis using the October issue

The control group for the September issue will be treated in October and we will test whether we observe similar effects for them. We do not expect that there are long run effects of the lower price for the September issue. Under that assumption, we can also use the early treated group as a comparison.

### Power

The sample will contain around 150 individuals, half assigned to treatment and half to control, assuming no or almost no sellers opt out of the intervention. We know from data from previous years that the lags have very strong predictive power for the outcome, amounting to an R-square of 0.71, so we consider that. At the conventional significance level of 0.05, an R-square of 0.7, and a power of 0.8, our sample size would allow for a minimum detectable effect of 0.25 standard deviations. Assuming only an R-squared of the controls of 0.5 and a sample of 130 individuals, the minimum detectable effect will be 0.35.

### Archive

The pre-analysis plan is archived before any post-intervention data are collected. We archive it at the registry for randomized controlled trials in economics held by The American Economic Association: https://www.socialscienceregistry.org/ on September 16, 2024.

### References

Agarwal, S., Diao, M., Pan, J., & Sing, T. F. (2015). Are Singaporean Cabdrivers Target Earners? Available at: http://dx.doi.org/10.2139/ssrn.2338476

Athey, S., & Imbens, G. W. (2017). The econometrics of randomized experiments. In Handbook of Economic Field Experiments (Vol. 1, pp. 73–140). Elsevier.

Belloni, A., Chernozhukov, V., & Hansen, C. (2014). Inference on treatment effects after selection among high-dimensional controls. The Review of Economic Studies, 81(2), 608–650.

Camerer, C., Babcock, L., Loewenstein, G., & Thaler, R. (1997). Labor supply of New York City cabdrivers: One day at a time. Quarterly Journal of Economics, 112, 407–441.

Duong, H. L., Chu, J., & Yao, D. (2022). Taxi drivers' response to cancellations and no-shows: New evidence for reference-dependent preferences. Management Science, 69(1), 179–199.

Farber, H. (2015). Why you can't find a taxi in the rain and other labor supply lessons from cabdrivers. Quarterly Journal of Economics, 130, 1975–2026.

Kahneman, D., & Tversky, A. (1979). Prospect theory: An analysis of decision under risk. Econometrica, 47(2), 1325–1348.

Leeson, P., Hardy, R. A., & Suarez, P. A. (2022). Hobo Economicus. Economic Journal, 132, 2325–2338.

6



\end{document}